\documentclass[10pt,prb,aps,twocolumn,showpacs,floatfix,longbibliography]{revtex4-1}
\usepackage[latin9]{inputenc}
\usepackage{array}
\usepackage{amsmath}
\usepackage{graphicx}

\makeatletter

\providecommand{\tabularnewline}{\\}
\usepackage{tikz}
\usetikzlibrary{calc}

\usepackage{epstopdf}
\usepackage{verbatim}
\usepackage{color}
\usepackage{subfigure}
\usepackage{tabularx}
\usepackage{amsfonts}
\usepackage{wasysym}
\usepackage{bm}

\newcommand{\ket}[1]{\ensuremath{|#1\rangle}}

\newcommand{\bs}{\boldsymbol}

\newcolumntype{C}[1]{>{\centering\let\newline\\\arraybackslash\hspace{0pt}}m{#1}}

\makeatother

\begin{document}

\title{Selective measurements of intertwined multipolar orders:\\ non-Kramers doublets on a triangular lattice}

\author{Chang-Le Liu$^{1}$}
\thanks{These authors contributed equally.}
\author{Yao-Dong Li$^{1,2}$}
\thanks{These authors contributed equally.}
\author{Gang Chen$^{1,3,4}$}
\email{gangchen.physics@gmail.com}

\affiliation{$^{1}$State Key Laboratory of Surface Physics and Department of Physics, Fudan University,
Shanghai 200433, China}
\affiliation{$^{2}$Department of Physics, University of California Santa Barbara, Santa Barbara, CA, 93106, USA}
\affiliation{$^{3}$Center for Field Theory and Particle Physics, Fudan University, Shanghai, 200433, China}
\affiliation{$^{4}$Collaborative Innovation Center of Advanced Microstructures, Nanjing University, Nanjing, 210093, China}

\date{\today}
\begin{abstract}
Motivated by the rapid experimental progress on the spin-orbit-coupled Mott insulators,
we propose and study a generic spin model that describes the interaction between the
non-Kramers doublets on a triangular lattice and is relevant for triangular lattice
rare-earth magnets. We predict that the system supports both pure quadrupolar orders
and intertwined multipolar orders in the phase diagram. Besides the multipolar orders,
we explore the magnetic excitations to reveal the dynamic properties of the systems.
Due to the peculiar properties of the non-Kramers doublets and the selective coupling to the
magnetic field, we further study the magnetization process of the system in the magnetic
field. We point out {\sl the selective measurements} of the static and dynamic properties
of the intertwined multipolarness in the neutron scattering, NMR and $\mu$SR probes
and predict the experimental consequences. The relevance to the existing materials such as
TmMgGaO$_4$, Pr-based and Tb-based magnets, and many ternary chalcogenides is discussed.
Our results not only illustrate the rich physics and the promising direction in the interplay
between strong spin-orbit-entangled multipole moments and the geometrical frustration,
but also provide a general idea to use non-commutative observables to reveal the dynamics of the
hidden orders. 
\end{abstract}
\maketitle


\section{Introduction}
\label{sec1}


There has been an intensive activity and interest in correlated matters with strong spin-orbit coupling~\cite{WCKB}. Various interesting quantum phases have been proposed, and the emergence of these rich phases is impossible in the absence of the strong spin-orbit coupling. More substantially, the abundance of candidate materials allows a rapid experimental progress of this field. In fact, the physical models for many relevant physical systems have not yet been constructed and thus not been explored carefully. This requires the knowledge of the microscopic nature of the relevant degrees of freedom. To establish the connection with the experimental observables, one needs further to understand the appearance of the physical properties for different phases of these newly constructed models. In this work, we carry out these thoughts and study the spin-orbit-coupled Mott insulators with non-Kramers doublets on a triangular lattice.

Since the discovery and the proposal of the spin liquid candidate material YbMgGaO$_{4}$~\cite{srep,YueshengPRL,Yaodong2016PRB,YaoShenNature,Martin2016,PhysRevB.97.125105,PhysRevB.96.054445,YueshengmuSR}, the triangular lattice rare-earth magnets have received more attention recently~\cite{Yaodong2016PRB,Yaodong2016PRB2,ChangPRB,PhysRevB.95.165110,PhysRevLett.120.087201,Parker2018,Shiyan2016,Toth1705,Yuesheng1704,yuesheng1702,PhysRevB.95.165110,JunZhaounpub,SciPostPhys.4.1.003,PhysRevLett.119.157201,PhysRevLett.120.037204,1710.06860,1801.01130}. Many isostructural rare-earth magnets such as RCd$_{3}$P$_{3}$, RZn$_{3}$P$_{3}$, RCd$_{3}$As$_{3}$, RZn$_{3}$As$_{3}$ \cite{triangle1, triangle2, triangle3}, KBaR(BO$_{3}$)$_{2}$ ~\cite{triangle4} (R is a rare-earth atom), and many ternary chalcogenides~\cite{OHTANI,Sato} are now proposed. In these systems, the rare-earth atoms form a perfect triangular lattice. The combination of the spin-orbit coupling of the 4f electrons and the crystal electric field creates a local ground state doublet that is described by an effective spin-1/2 local moment at each rare-earth site. These rare-earth local moments then interact with each other and describe the low-temperature magnetic properties of the system. In most cases, the superexchange interaction is short-ranged, and nearest-neighbor exchange interaction with further neighbor dipolar interaction is sufficient due to the strong spatial localization of the 4f electron wavefunction. These materials provide a natural setting to study {\sl the interplay between strong spin-orbit entanglement and geometrical frustration} in both theory and experiments.

\begin{figure}[b]
\includegraphics[width=8cm]{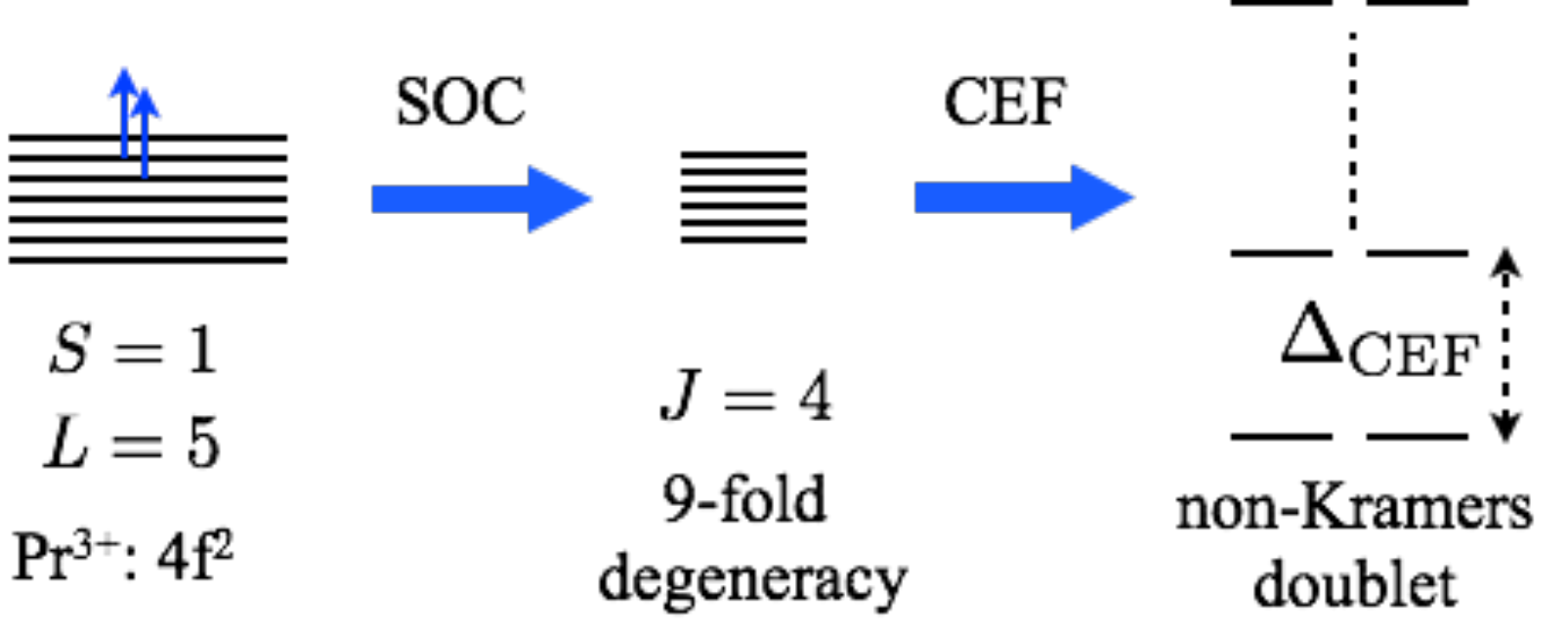}
\caption{The combination of spin-orbit coupling and the $D_{3d}$ crystal electric
field generates a non-Kramers doublet ground state for the Pr$^{3+}$ ion.
Here ``SOC'' refers to spin-orbit coupling, and ``CEF'' refers to crystal electric field.
Other ions such as Tm$^{3+}$ and Tb$^{3+}$ could potentially support non-Kramers doublets.}
\label{fig1}
\end{figure}

In the list of relevant physical models for the rare-earth triangular lattice magnets, we have explored the usual Kramers doublets and the dipole-octupole doublets~\cite{PhysRevLett.112.167203,PhysRevB.95.041106} in the previous works~\cite{Yaodong2016PRB,Yaodong2016PRB2}. In particular, the anisotropic spin model~\cite{Yaodong2016PRB,PhysRevB.97.125105} for the usual Kramers doublets was suggested to be relevant for the spin liquid candidate YbMgGaO$_{4}$ and many other rare-earth triangular lattice magnets with Kramers ion. In this work, we turn our attention to the non-Kramers doublet on the triangular lattice that has been advocated in the end of Ref.~\onlinecite{Yaodong2016PRB}, and complete the full list of the microscopic spin models for the triangular lattice rare-earth magnets. Unlike the usual Kramers doublets, the mixed multipolar natures of spin components for the non-Kramers doublets greatly simplify the spin Hamiltonian. For the non-Kramers doublets~\cite{PhysRevB.86.104412,PhysRevLett.105.047201,Kivelson}, the longitudinal spin component behaves as the magnetic dipole moment, while the transverse spin components behave as the magnetic quadrupole moment. Therefore, the time reversal symmetry and the hermiticity condition forbid the coupling between the longitudinal and the transverse components. Moreover, the ordering in the longitudinal spin components and the ordering in the transverse components have to be distinct and necessarily correspond to different phases and phase transitions. The purpose of this paper is to understand the intertwined multipolar ordering structures and the relevant experimental phenomena for the non-Kramers doublets on the triangular lattice.

The magnetic dipolar order can be directly visible through the conventional magnetic measurements such as the NMR and neutron diffraction experiments. The magnetic quadrupolar order (or equivalently, spin nematicity) preserves the time reversal symmetry and is often not quite visible in such conventional measurements. However, the dipole component, that is orthogonal to the quadrupole component, could then create quantum fluctuations for the quadrupole component and lead to coherent spin wave excitations. This {\sl orthogonal effect} allows the detection of the spin wave spectra via the inelastic neutron scattering measurements. If the quadrupolar order breaks the translation symmetry and enlarges the unit cell, the symmetry breaking pattern may not be quite visible in the static measurement, but is clear in the dynamic measurements. Thus, we study the magnetic excitations in the multipolar ordered phases. We establish the key connections between the underlying multipolar structure and the features in the excitation spectra. The orthogonal effect of the dipole component on the quadrupole component further lies in the coupling to the external magnetic field. The magnetic field only couples linearly to the dipole component, and thus, the magnetization and the magnetic susceptibility indirectly suggest the underlying quadrupolar order and transition.

The following part of the paper is organized as follows. In Sec.~\ref{sec2}, we propose the relevant physical model for the non-Kramers doublets on a triangular lattice and explain the physical significance of the spin operators. In Sec.~\ref{sec3}, we employ several different methods to obtain the full phase diagram of this model. Since many states have an emergent $U(1)$ symmetry at the mean-field level, in Sec.~\ref{sec4}, we study the quantum order by disorder phenomena for two representative states on our phase diagram. In Sec.~\ref{sec5}, we study the dynamic properties of the distinct phases that can serve as the experimental probes of the underlying multipolar orders. In Sec.~\ref{sec6}, we point out the unique magnetization process due to the selective coupling of the moments to the external magnetic field. Finally in Sec.~\ref{sec7}, we discuss the experimental detection of various phases and summarize with a materials' survey. In Appendix.~\ref{appendix1}, we explain the relevance of the model to the Kitaev interaction.
In Appendix.~\ref{appendix2}, we give the explanation of the non-Kramers doublet for the 
case of the spin-1 moments. 
In Appendix.~\ref{appendix3}, we show the complete spin wave dispersions for different phases.

\begin{figure}[t]
\includegraphics[width=3.6cm]{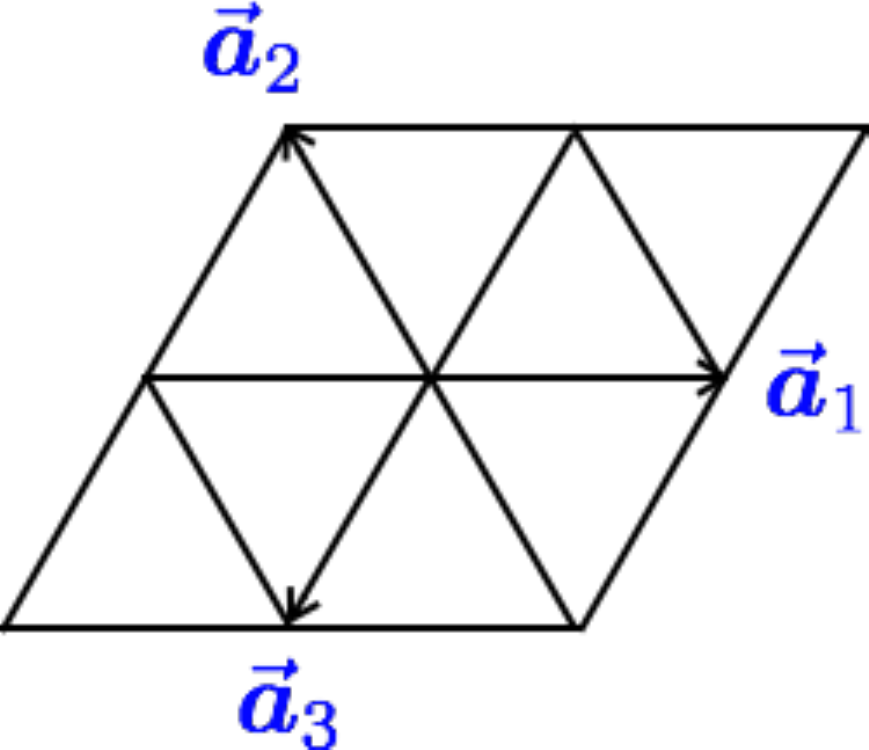}
\hspace{0.6cm}
\includegraphics[width=3.2cm]{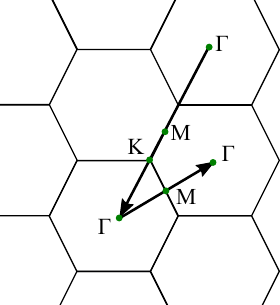}
\caption{(a) The triangular lattice with three distinct neighboring bonds and interactions.
The phase parameter $\gamma_{ij}$ depends on the bond orientation, which reflects the
spin-orbit-entangled nature of the local moments.
(b) The definition of the Brillouin zone for the triangular lattice.}
\label{fig2}
\end{figure}

\section{Model Hamiltonian}
\label{sec2}

\begin{table*}[ht]
\begin{tabular}{lp{10cm}c}
\hline\hline
Local doublets  & The nearest-neighbor spin Hamiltonians on the triangular lattice  & Reference \tabularnewline \hline
Usual Kramers doublet  & $H=\sum_{\langle ij\rangle}J_{zz}S_{i}^{z}S_{j}^{z}+J_{\pm}(S_{i}^{+}S_{j}^{-}+S_{i}^{-}S_{j}^{+})+J_{\pm\pm}(\gamma_{ij}S_{i}^{+}S_{j}^{+}+\gamma_{ij}^{\ast}S_{i}^{-}S_{j}^{-})$
$-\frac{iJ_{z\pm}}{2}[(\gamma_{ij}^{\ast}S_{i}^{+}-\gamma_{ij}S_{i}^{-})S_{j}^{z}+S_{i}^{z}(\gamma_{ij}^{\ast}S_{j}^{+}-\gamma_{ij}S_{j}^{-})]$  & Refs.~\onlinecite{Yaodong2016PRB,YueshengPRL} \tabularnewline
Dipole-octupole doublet  & $H=\sum_{\langle ij\rangle}J_{z}S_{i}^{z}S_{j}^{z}+J_{x}S_{i}^{x}S_{j}^{x}+J_{y}S_{i}^{y}S_{j}^{y}+J_{yz}(S_{i}^{z}S_{j}^{y}+S_{i}^{y}S_{j}^{z})$  & Ref.~\onlinecite{Yaodong2016PRB2} \tabularnewline
Non-Kramers doublet  & $H=\sum_{\langle ij\rangle}J_{zz}S_{i}^{z}S_{j}^{z}+J_{\pm}(S_{i}^{+}S_{j}^{-}+S_{i}^{-}S_{j}^{+})+J_{\pm\pm}(\gamma_{ij}S_{i}^{+}S_{j}^{+}+\gamma_{ij}^{\ast}S_{i}^{-}S_{j}^{-})$  & This work \tabularnewline
\hline\hline
\end{tabular}\caption{The relevant spin Hamiltonians for three different doublets on the
triangular lattice. The models for the usual Kramers doublet and the
dipole-octupole doublet have been obtained in the previous works. }
\label{tab1}
\end{table*}

Apart from YbMgGaO$_{4}$, RCd$_{3}$P$_{3}$, RZn$_{3}$P$_{3}$, RCd$_{3}$As$_{3}$, RZn$_{3}$As$_{3}$, KBaR(BO$_{3}$)$_{2}$, and many ternary chalcogenides (LiRSe$_2$, NaRS$_2$, NaRSe$_2$, KRS$_2$, KRSe$_2$, KRTe$_2$, RbRS$_2$, RbRSe$_2$, RbRTe$_2$, CsRS$_2$, CsRSe$_2$, CsRTe$_2$, etc) are known to have the rare-earth local moments on the triangular lattices, where R is the rare-earth atom. These chemical properties of the rare-earth atoms are quite similar, and thus it is often possible to substitute one for the other. The rare-earth ion such as Yb$^{3+}$ and Sm$^{3+}$, that contains odd number of electrons, is the Kramers' ion and forms a ground state doublet whose two fold degeneracy is protected by the time reversal symmetry and the Kramers' theorem. The non-Kramers ion like Pr$^{3+}$ and Tb$^{3+}$ contains an even number of electrons per site (see Fig.~\ref{fig1}). The spin-orbit coupling of the 4f electrons entangles the total spin moment and the orbital angular momentum, and leads to a total moment $J$ that is an integer. The crystal electric field then splits the $2J+1$ fold degeneracy and sometimes leads to a two-fold degenerate ground state doublet. Although both Kramers doublet and non-Kramers doublet are two-dimensional irreducible representation of the point group, the two-fold degeneracy of the Kramers doublets is further protected by the time reversal symmetry, and the degeneracy of the non-Kramers doublets is merely protected by the lattice symmetry. For these non-Kramers doublet, one then introduces an effective spin-1/2 operator, ${\boldsymbol{S}}_{i}$, that acts on the two-fold degenerate ground state doublet at each lattice site
(see Appendix.~\ref{appendix2} for a more detailed discussion for a specific case.)

Although the effective spin-1/2 operator is introduced to describe
the non-Kramers doublet, the actual wavefunctions of the non-Kramers
doublets are still integer spins in nature. As a result, the transformation
of these effective spin-1/2 operators for the non-Kramers doublet
is quite different from the effective spin-1/2 operators for the Kramers
doublet under the time reversal symmetry. Specifically, the longitudinal
component, $S_{i}^{z}$, is odd under time reversal and transforms
as a magnetic dipole moment, and the transverse components, $S_{i}^{x}$
and $S_{i}^{y}$, are even under time reversal and transform as the
magnetic quadrupolar moment. Therefore, the generic symmetry-allowed
spin Hamiltonian, that describes the interaction between the non-Kramers
doublets on the triangular lattice, is simpler than the one for the
Kramers doublets and is given as~\cite{Yaodong2016PRB}
\begin{eqnarray}
H & = & \sum_{\langle ij\rangle}J_{zz}S_{i}^{z}S_{j}^{z}+J_{\pm}(S_{i}^{+}S_{j}^{-}+S_{i}^{-}S_{j}^{+})\nonumber \\
 &  & \quad\quad+J_{\pm\pm}(\gamma_{ij}S_{i}^{+}S_{j}^{+}+\gamma_{ij}^{\ast}S_{i}^{-}S_{j}^{-}),
\label{eq1}
\end{eqnarray}
in which, $\gamma_{ij}$ is a bond-dependent phase factor, and takes
$1$, $e^{i2\pi/3}$ and $e^{-i2\pi/3}$ on the ${\boldsymbol{a}}_{1},{\boldsymbol{a}}_{2}$
and ${\boldsymbol{a}}_{3}$ bond (see Fig.~\ref{fig2}), respectively.
As shown in Tab.~\ref{tab1}, this model differs from the Kramers
doublet model by the absence of the coupling between the transverse
components and the longitudinal component.

Besides the standard expression of the model in Eq.~(\ref{eq1}), in the Appendix~\ref{appendix1}
we further recast the model into a different form where the Kitaev interaction is explicitly shown.

\section{Phase diagram}
\label{sec3}

In this section, we carry out several complementary approaches to determine the
classical or mean-field phase diagram of the spin model defined in Eq.~(\ref{eq1}).
The model is apparently frustrated, and a complicated phase diagram is expected.

We first notice that in the model, the spin rotation around the $z$ direction by
$\pi/4$ transforms $S^{\pm}\rightarrow\mp iS^{\pm}$, and the couplings in the model
transform as
\begin{eqnarray}
{J_{zz}\rightarrow J_{zz}},\quad
{J_{\pm}\rightarrow J_{\pm}},\quad
{J_{\pm\pm}\rightarrow-J_{\pm\pm}}.
\end{eqnarray}
Therefore, we can focus on the ${J_{\pm\pm}>0}$ region of the phase diagram.
Moreover, as most of relevant materials are antiferromagnets, we choose
${J_{zz}>0}$ in our analysis for the reason that will be clear later.
Our results are summarized in Fig.~\ref{fig3} and Tab.~\ref{tab2}.

\subsection{Pure quadrupolar orders}

To start with, we tackle this model in the spirit of a Weiss-type mean-field approach.
This approach is qualitively correct if the ground state of the spin model supports
long-range orders with local on-site order parameters. This approach often provides
some very basic information about the ground state properties of the system. Within
this approach, we treat the spin as a classical vector and optimize the energy by
choosing a proper spin configuration. The classical spin vector is subjected to a
local constraint ${|{\boldsymbol{S}}_{i}|=S}$, and is thus often difficult to deal
with. One can nevertheless try to solve for the ground state of the mean-field Hamiltonian
with a relaxed global constraint, ${\sum_{i}{\boldsymbol{S}}_{i}^{2}=NS^{2}}$,
where $N$ is total number of spins, which does not necessarily respect the local
spin constraint. When it does, this state is the actual classical ground state of
the classical spin Hamiltonian. This method is often known as ``Luttinger-Tisza''
method~\cite{LuttingerTisza}.

\begin{figure}[thp]
\centering
\includegraphics[width=8.4cm]{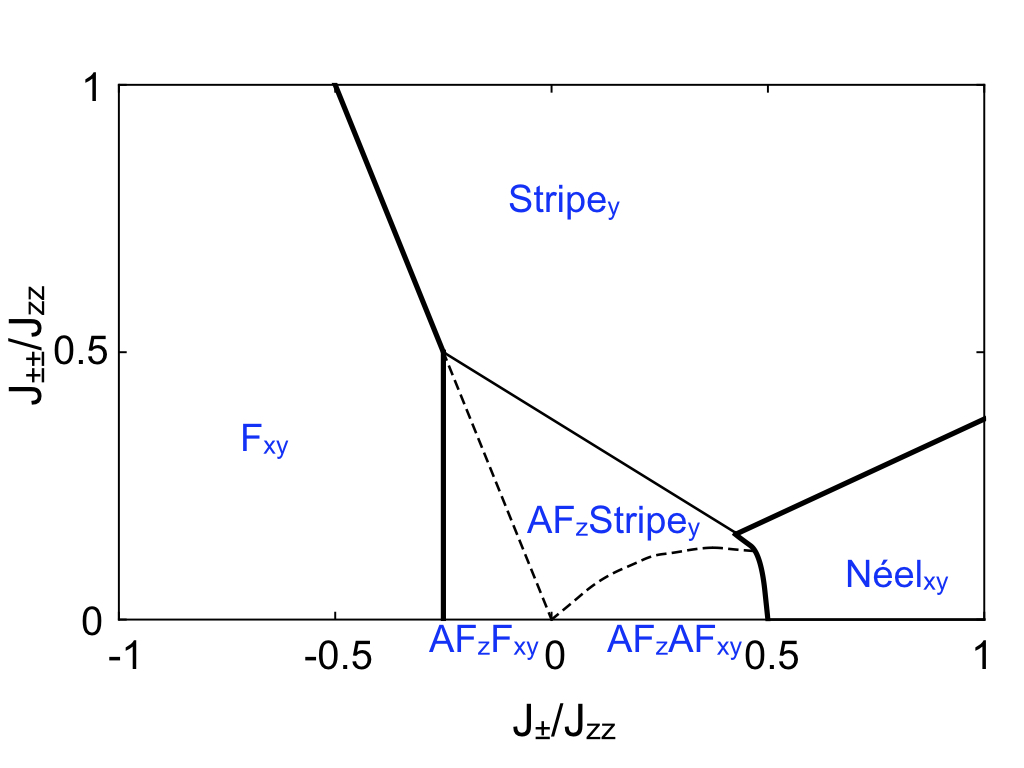}
\caption{The mean-field ground state phase diagram of the model in Eq.~(\ref{eq1}) with ${J_{zz} > 0}$. We find the Stripe$_{\text y}$ state for large $J_{\pm\pm}$ regardless of the sign of $J_\pm$, the F$_{\text{xy}}$ state for large negative $J_\pm$, and the N\'{e}el state for a large and positive $J_\pm$. Thick curves refer to first order transitions, and thinner curves refer to second order transitions. The dashed phase boundaries are determined by comparing the energy of AF$_{\text z}$Stripe$_{\text y}$ states with those of AF$_{\text z}$F$_{\text{xy}}$ and
AF$_{\text z}$AF$_{\text{xy}}$. The transitions across the dashed lines are complicated and may involve other competing  states that is not well captured by our mean-field approach. The spin configurations of all ordered states are illustrated in Fig.~\ref{fig4}.}
\label{fig3}
\end{figure}

\begin{figure}[b]
\centering
\includegraphics[width=8.4cm]{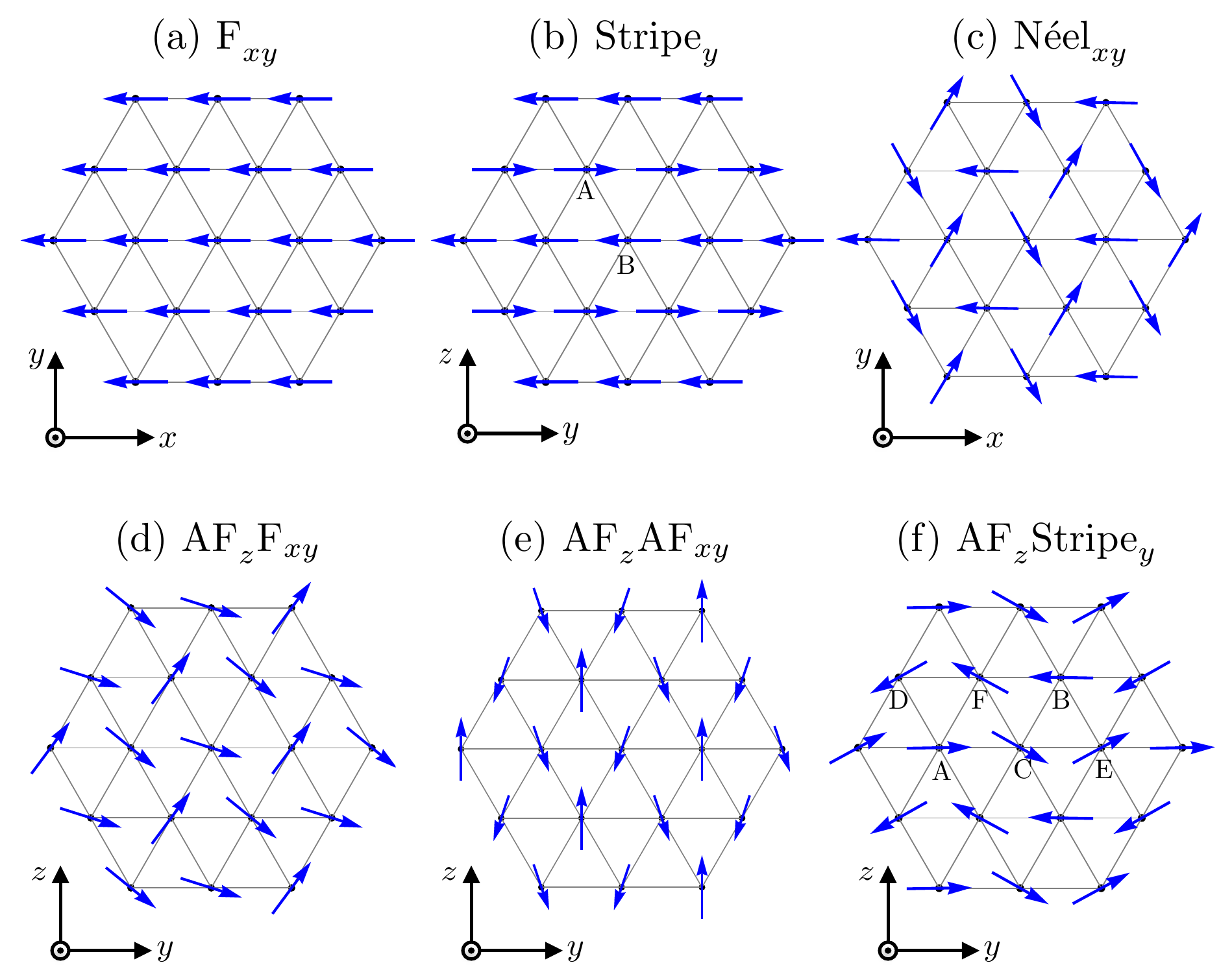}
\caption{Real-space spin configurations of the mean-field ground states found in Fig.~\ref{fig3}.
(a) The ferromagnetic quadrupolar order with spins aligned in xy-plane, which we name the F$_{\text{xy}}$ order.
There is a global $U(1)$ degeneracy in the xy-plane.
(b) The antiferromagnetic quadrupolar stripe order with spins aligned in $y$-direction, which we dub ``Stripe$_{\text y}$''.
(c), (d), The AF$_{\text z}$F$_{\text{xy}}$ and AF$_{\text z}$AF$_{\text{xy}}$ orders in Fig.~\ref{fig3}
for small $J_{\pm\pm}$ and small $J_{\pm}$. Both orders have a 3-sublattice structure, consistent with results from
Refs.~\cite{troyer2005supersolid, balents2005supersolid, danshita2014xxz}. As in (b), the component in xy-plane has a global $U(1)$ degeneracy for reasons explained in the text. (e) The $120^\circ$-N\'{e}el order stabilized by a large positive $J_{\pm}$. In all figures, we draw the coordinate system of the spin space to help visualization. The coordinate system of the real space always takes the same convention in Fig.~\ref{fig2}.
}
\label{fig4}
\end{figure}

In most parts of the phase diagram, the Luttinger-Tisza method can correctly reproduce the classical ground state. In the regimes where the method fails, we adopt a multi-sublattice mean-field ansatz to minimize the ground state energy. This approach is obviously simplified and cannot capture some of more complicated magnetic orders or absense of magnetic orders due to strong frustration. 
The phase diagram obtained from mean-field theory is shown in Fig.~\ref{fig3}. We found a family of long-range ordered phases as illustrated in Fig.~\ref{fig4}. When one of the couplings is dominant, the frustration is suppressed, and the Luttinger-Tisza method works out well. This is the regime in which either $J_\pm$ or $J_{\pm\pm}$ is dominant, and we have
\begin{itemize}
\item F$_{\text{xy}}$ state when $J_{\pm}$ is large and negative. The ordering wavevector is at $\Gamma$ point. In this state, the quadrupole components $S^{x}$ and $S^{y}$ align in the same direction in xy-plane. At the mean-field level, this state has an emergent $U(1)$ degeneracy under the global rotation of an arbitrary angle about $S^z$. This is a bit surprising here since the microscopic model only has a discrete lattice symmetry due to the spin-orbit coupling. Thus, the emergent continuous degeneracy here and below is completely accidental, and quantum fluctuation beyond the mean-field theory should lift this degeneracy. This is the well-known order by quantum disorder. Moreover, due to this emergent continuous degeneracy, the excitation spectrum with respect to the quadrupolar order would have a pseudo-Goldstone mode that is nearly gapless. In Sec.~\ref{sec4}, we carry out an explicit calculation to discuss this order by quantum disorder in this regime.
\item 120$^{\circ}$ N\'{e}el state with pure quadrupolar orders appears as the ground-state in the large $J_{\pm}$ regime. In this state spins lie in the xy plane and each spins are arranged 120$^{\circ}$ to its nearest neighbor, thus the ordering wavevector occurs at the $K$ point. The state has non-vanishing quadrupolar components $S^{x}$ and $S^{y}$. We find that this state has degenerate energies under effective spin rotation of arbitrary angle about $S^z$, so this state has emergent $U(1)$ degeneracy. For the same reason as the F$_{\text{xy}}$ state, there would be a pseudo-Goldstone mode
at $\Gamma$ point.
\item Stripe$_{\text y}$ order when $J_{\pm\pm}$ is large. In this state, the quadrupolar component $S^{y}$ is aligned in alternating directions for alternating rows of spins. The ordering wavevector is at $M$ point. The spin-wave excitation is in general fully gapped.
\end{itemize}

The N\'{e}el and F$_{\text{xy}}$ orders can be understood in the XXZ limit, where a large antiferromagnetic $J_\pm$ induces the N\'{e}el order with the 3-sublattice structure, and a large ferromagnetic $J_\pm$ stabilizes the ferromagnetic order. The somewhat surprising emergent $U(1)$ symmetry is due to the cancelling $\gamma_{ij}$ phase factors of the anisotropic spin coupling term $J_{\pm\pm}$. The above three phases are purely quadrupolar orders, and are completely hidden in the magnetization measurements. Since they are absent of dipolar orders, even the elastic neutron scattering measurement cannot resolve these states. The dipolar spin component $S^z$, however, can create a coherent spin-wave excitation with respect to the quadrupolar ordered phases. Thus, despite the seemingly absence in the conventional magnetization measurements, the quadrupolar orders can nevertheless be detected via the inelastic neutron scattering experiments. We will explore this in Sec.~\ref{sec4}.

\subsection{Intertwined multipolar orders}

Next we focus on the case with dominant $J_{zz}$ that is presumably the most frustrated regime and thus supports strong quantum fluctuations. We here implement a traditional self-consistent Weiss mean-field theory by replacing the generic pair-wise spin interactions as
\begin{eqnarray}
S^{\mu}_i S^{\nu}_j \rightarrow \langle S^{\mu}_i \rangle S^{\nu}_j + S^{\mu}_i \langle S^{\nu}_j \rangle
- \langle S^{\mu}_i \rangle \langle S^{\nu}_j \rangle ,
\end{eqnarray}
where $\langle S^{\mu}_i \rangle$ is the order parameter of the mean-field state and should be solved self-consistently. For this purpose, one first needs to set up a mean-field ansatz for the order parameters. From the experience of the XXZ model, one should at least choose a 3-sublattice mean-field ansatz. Here, to be a bit more general, we choose a 6-sublattice mean-field ansatz for some parameter regime. The local stability of the ground state is examined
by the spin wave calculation. If the mean-field ground state is locally unstable, the spin wave spectra will
no longer be real and positive.

Within this self-consistent mean-field approach, we find three types of intertwined multipolar long-range orders that are depicted in Fig.~\ref{fig3} and Fig.~\ref{fig4} and listed below,
\begin{itemize}
\item AF$_{\text z}$F$_{\text{xy}}$ for negative $J_\pm$ and small $J_{\pm\pm}$. In this state the spins have both nonzero antiferromagnetically ordered dipolar $S^z$ and ferromagnetically ordered quadrupolar $S^{x,y}$ components. There is an emergent $U(1)$ degeneracy generated by the spin rotation about tne $S^z$ direction.
\item AF$_{\text z}$AF$_{\text{xy}}$ for positive $J_\pm$ and small $J_{\pm\pm}$. In this state the spins have both nonzero antiferromagnetically ordered dipolar $S^z$ and antiferromagnetically ordered quadrupolar $S^{x,y}$ components. This state also has the emergent $U(1)$ degeneracy in the xy plane of the spin space.
\item AF$_{\text z}$Stripe$_{\text y}$ at larger $J_{\pm\pm}$. This phase is found proximate to Stripe$_{\text y}$ phase via the second-order transition. It has a similar pattern with the Stripe$_{\text y}$ state where the quadrupolar moment  $S^y$ orders in the stripe-like pattern. It also develops magnetic order in the dipole component $S^z$.
\end{itemize}

All the above states carry intertwined multipolar orders, supporting both dipolar and quadrupolar orders. Here we provide the physical understanding for the emergence of these interesting orders. The AF$_{\text z}$F$_{\text{xy}}$ and AF$_{\text z}$AF$_{\text{xy}}$ states are found to be the exact ground states in the XXZ limit where ${J_{\pm\pm} = 0}$, and are known as the supersolid orders in this limit, for which both the ``boson density'' $S^z$ and the ``superfluid order parameters'' $S^{x, y}$ are non-vanishing. These supersolid orders are no longer the exact ground states for small values of $J_{\pm\pm}$. Moreover, the notion of ``supersolidity'' is ill-defined because the $J_{\pm\pm}$ interaction explicitly breaks the $U(1)$ spin rotational symmetry of the XXZ model. In fact, with a small $J_{\pm\pm}$ near the XXZ limit, the AF$_{\text z}$F$_{\text{xy}}$ and AF$_{\text z}$AF$_{\text{xy}}$ states become unstable from our linear spin wave calculation and may turn into some incommensurate states. The incommensurate states are not well captured by our self-consistent mean-field approach that assumes commensurate states from the starting point. In the phase diagram we nevertheless label the small $J_{\pm\pm}$ regime with the supersolid orders (AF$_{\text z}$F$_{\text{xy}}$ and AF$_{\text z}$AF$_{\text{xy}}$ states).

\begin{table}[t]
\begin{tabular}{c|c|c}
\hline\hline
states  & order types & elastic neutron \\
\hline
F$_{\text{xy}}$ & pure quadrupolar & no Bragg peak  \\
\hline
120$^{\circ}$ N\'{e}el & pure quadrupolar & no Bragg peak  \\
\hline
Stripe$_{\text y}$ & pure quadrupolar & no Bragg peak  \\
\hline
AF$_{\text z}$F$_{\text{xy}}$ & intertwined multipolar & Bragg peak at K\\
\hline
AF$_{\text z}$AF$_{\text{xy}}$ & intertwined multipolar & Bragg peak at K\\
\hline
AF$_{\text z}$Stripe$_{\text y}$ & intertwined multipolar & Bragg peak at K\\
\hline\hline
\end{tabular}\caption{The list of ordered phases in the phase diagram of Fig.~\ref{fig3}.}
\label{tab2}
\end{table}

The AF$_{\text z}$Stripe$_{\text y}$ state has intertwined dipolar $S^z$ order and quadrupolar
$S^y$ order that result from the competition between $J_{\pm\pm}$ and $J_{zz}$.
In the Ising limit with ${J_{\pm\pm} \ll J_{zz}}$, it is well-known that the ground state manifold
is extensively degenerate: the energy of a state is minimized as long as in each triangle Ising spins
are not simultaneously parallel to each other. In the supersolid orders, this is manifested in the
spin pattern where the signs of the $S^z$ component is $(+,-,-)$ or $(+,+,-)$ in each triangle.
Away from the Ising limit, a nonzero $J_{\pm\pm}$ allows the system to fluctuate within
the extensively degenerate manifold of Ising spins, and therefore lifts the extensive
degeneracy. This is quite analogous to the effect of the transverse field on top of the
antiferromagnetic Ising interaction on the triangular lattice. The ground state in our
case is such that the quadrupolar $S^y$ component is maximized and ordered in a stripe-like
pattern to optimize the $J_{\pm\pm}$ term, while the dipolar $S^z$ component orders in
such a pattern where the signs of the $S^z$ component is $(+,-,0)$ in each triangle.
As we show in Fig.~\ref{fig4}, the combined structure of the dipolar and quadrupolar
orders has a 6-sublattice structure.

Unlike the pure quadrupolar order in the previous subsection, the intertwined multipolar orders are not completely invisible in the conventional magnetic measurement. The multiple-sublattice structure of the dipolar components can be detected through the usual bulk magnetization measurements such as NMR, $\mu$SR, and elastic neutron scattering measurements. Again, the quadrupolar orders hide themselves from such measurements. Thus, the intertwined multipolarness is only partially visible.

Here, the presence of the intertwined multipolar order in this part of the phase diagram results from
{\sl the combination of the geometrical frustration and the multipolar nature} of the local moment.
With only geometry frustration, the system would simply support the conventional magnetic orders.
With only spin-orbit-entangled local moments and the multipolar structure of the
local moment, the system would not give an {\sl intertwined} multipolar ordering structure.
It is {\sl the combination of the geometrical frustration and the multipolar nature} of the
local moment that gives rise to the intertwined multipolar ordering structure.

\section{Quantum order by disorder}
\label{sec4}

\begin{figure}
\includegraphics[width=6cm]{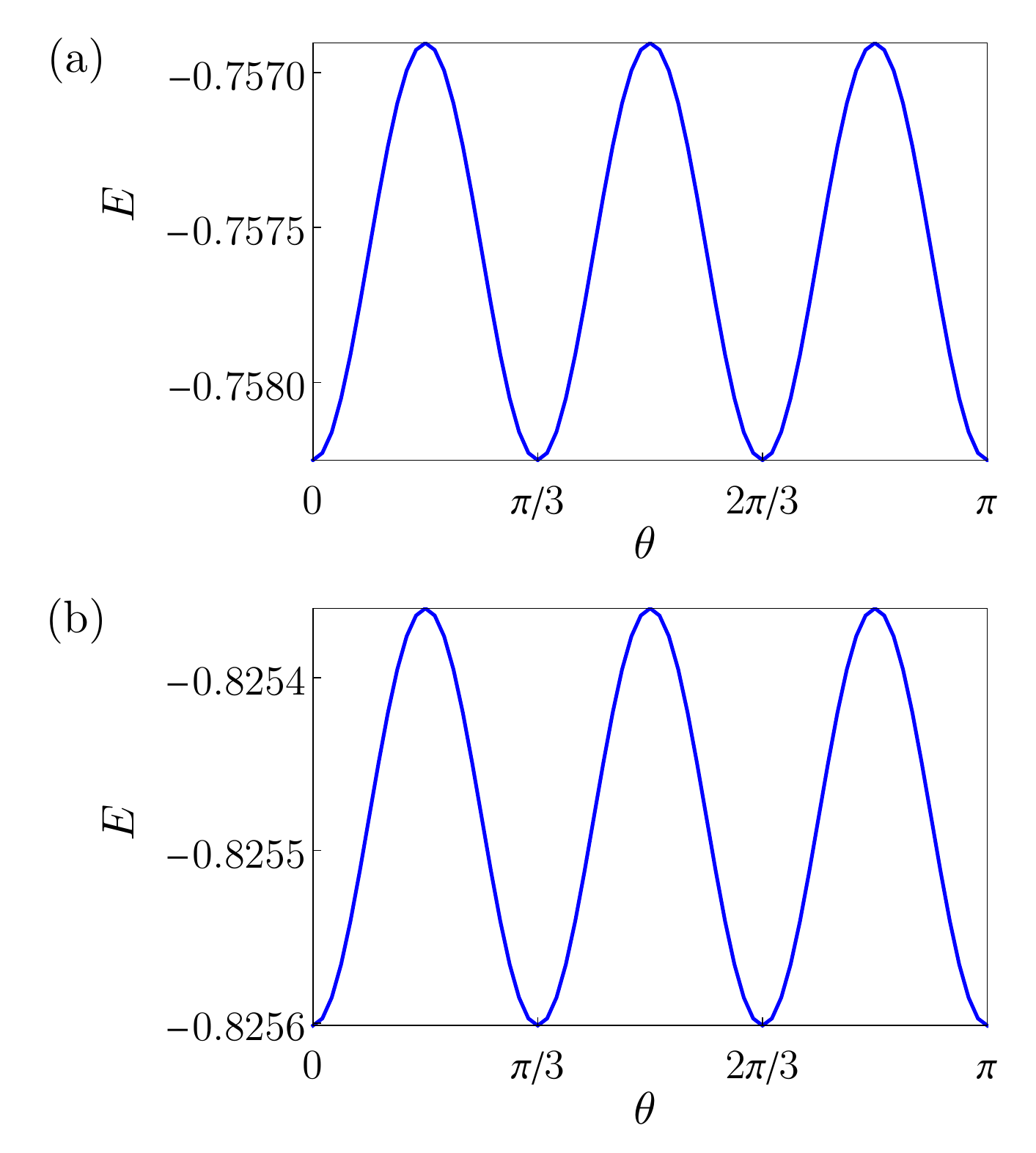}
\caption{Energy per spin taking into account quantum zero-point energy vs.
the azimuth angle $\theta$ of spins for (a) the F$_{\text{xy}}$
state and (b) the 120$^{\circ}$ N\'eel$_{\text{xy}}$ state. Here
we take the parameter ${J_{\pm}=0.4J_{zz}}$, ${J_{\pm\pm}=0.4J_{zz}}$
for the F$_{\text{xy}}$ state and parameter ${J_{\pm}=0.9J_{zz}}$,
${J_{\pm\pm}=0.2J_{zz}}$ for the N\'eel$_{\text{xy}}$ state. 
The zero-point energy is calculated within the linear spin 
wave method.}
\label{fig:zeropoint}
\end{figure}

As we describe in previous sections, the system has only discrete spin-rotational symmetries,
thus it is a bit counter-intuitive that all phases except for Stripe$_{\text y}$ phase in the mean-field
phase diagram host emergent continuous $U(1)$ degeneracies in the xy plane of the spin space.
These continuous degeneracies are due to non-trivial bond-dependent $\gamma_{ij}$ phase factors
in the $J_{\pm\pm}$ interactions. It should be noted that these continuous degeneracies are
presented only at mean-field level, and in general should be lifted by quantum fluctuations.
Here we study the quantum fluctuation in the F$_{\text{xy}}$ state as an example.
In the F$_{\text{xy}}$ state, the mean-field state has spins align in xy-plane with an arbitrary
azimuth angle $\theta$ with respect to the x axis. If we take into account quantum fluctuations,
these degenerate states will have different zero-point energies so that the degeneracy is lifted.
This effect can be captured in the linear spin wave theory. For ${J_{\pm\pm}>0}$, it is shown
in the Fig.~\ref{fig:zeropoint} that the quantum fluctuation selects the ground state with
${\theta=n\pi/3}$ (${n\in{\mathbb{Z}}}$) such that the spins align along the bond
orientations in the F$_{\text{xy}}$ state. For the 120$^{\circ}$ N\'{e}el state,
similar results are obtained, and the spins are aligned along the bond orientations.
For other states with continuous degeneracies we expect similar degeneracy breaking.

Here we present our linear spin wave method that applies to multi-sublattice
configurations~\cite{petit2011,wallace1962,toth2015}. Let us assume that the
system has $M$-sublattice magnetic order. Each spin can be labeled by the
magnetic unit cell index ${\mathbf r}$ and sublattice index $s$. Assuming
spins with sublattice index $s$ has the direction pointing along the
unit vector $\mathbf{n}_{s}$, one can always associate two unit vectors
${\mathbf{u}_{s}\cdot\mathbf{n}_{s}=0}$ and ${\mathbf{v}_{s}=\mathbf{n}_{s}\times\mathbf{u}_{s}}$
so that $\mathbf{n}_{s}$, $\mathbf{u}_{s}$ and $\mathbf{v}_{s}$
are orthogonal with each other. Then we perform Holstein-Primakoff
transformation for the spin operator $\mathbf{S}_{{\mathbf r}s}$,
\begin{eqnarray}
\mathbf{n}_{s}\cdot\mathbf{S}_{{\mathbf r}s} & = & S-b_{{\mathbf r}s}^{\dagger}b^{}_{{\mathbf r}s},  \\
(\mathbf{u}_{s}+i\mathbf{v}_{s})\cdot\mathbf{S}_{{\mathbf r}s} & = &
({2S-b_{{\mathbf r}s}^{\dagger}b_{{\mathbf r}s}})^{ \frac{1}{2}}b_{{\mathbf r}s} , \\
(\mathbf{u}_{s}-i\mathbf{v}_{s})\cdot\mathbf{S}_{{\mathbf r}s} & = &
b_{{\mathbf r}s}^{\dagger}   ({2S-b_{{\mathbf r}s}^{\dagger}b_{{\mathbf r}s}})^{\frac{1}{2}}.
\end{eqnarray}
After performing Fourier transformation
\begin{equation}
b_{{\mathbf{r}}s}=\sqrt{\frac{M}{N}}\sum_{\mathbf{k}\in{\overline{\text{BZ}}}}b_{\mathbf{k}s}e^{i\mathbf{R}_{{\mathbf{r}}s}\cdot\mathbf{k}},
\end{equation}
the spin Hamiltonian can be rewritten in terms of boson bilinears as
\begin{equation}
H_{\text{sw}}=E_{0}+\frac{1}{2}\sum_{\mathbf{k}\in {\overline{\text{BZ}}}}
[\Psi(\mathbf{k}){}^{\dagger}h(\mathbf{k})\Psi(\mathbf{k})-\frac{1}{2}\mbox{tr}\,h(\mathbf{k})],
\end{equation}
where $E_{0}$ is the mean-field energy,
\begin{equation}
\Psi(\mathbf{k})=[b_{\mathbf{k}1}^{}, \cdots,b_{\mathbf{k}M}^{},b_{\mathbf{-k}1}^{\dagger},\cdots,b_{\mathbf{-k}M}^{\dagger}]^{T}
\end{equation}
and $h(\mathbf{k})$ is a ${2M \times 2M}$ Hermitian matrix, and
$\overline{\text{BZ}}$ is the magnetic Brillouin zone. $H_{\text{sw}}$ can
be diagonalized via a standard Bogoliubov transformation
${\Psi(\mathbf{k})=T_{\mathbf{k}}\Phi(\mathbf{k})}$ where
\begin{equation}
\Phi(\mathbf{k})= [ \beta_{\mathbf{k}1}^{},\cdots,
\beta_{\mathbf{k}M}^{},\beta_{\mathbf{-k}1}^{\dagger},
\cdots,\beta_{-\mathbf{k}M}^{\dagger} ]^{T},
\end{equation}
and ${T_{\mathbf{k}}\in SU(M,M)}$. Here ${SU(M,M)}$ refers to
indefinite special unitary group that is defined as~\cite{Goodman2009}:
\begin{equation}
SU(M,M)=\{ {g\in\mathbb{C}_{2M\times2M}}:g^{\dagger}\Sigma g=\Sigma,\det g=1\},
\end{equation}
where $\Sigma$ is the metric tensor and given as
\begin{equation}
\Sigma=\begin{pmatrix}I_{M\times M}&0\\
0 & -I_{M\times M}
\end{pmatrix}.
\end{equation}
It is straightforward to prove that such transformation preserves the boson commutation rules.
The diagonalized Hamiltonian reads
\begin{eqnarray}
H_{\text{sw}} & = & E_{0}+\frac{1}{2}\sum_{\mathbf{k}\in \overline{\text{BZ}}}[\Phi(\mathbf{k}){}^{\dagger}E(\mathbf{k})\Phi(\mathbf{k})-\frac{1}{2}\mbox{tr}\,h(\mathbf{k})]\nonumber \\
 & = & E_{0}+E_{r}+\sum_{\mathbf{k}\in \overline{\text{BZ}}}\omega_{\mathbf{k}s}\beta_{\mathbf{k}s}^{\dagger}\beta_{\mathbf{k}s}^{},
\end{eqnarray}
where $E(\mathbf{k})=\mbox{diag}[\omega_{\mathbf{k}1},\cdots,\omega_{\mathbf{k}M},\omega_{-\mathbf{k}1},\cdots,\omega_{-\mathbf{k}M}]$
and
\begin{equation}
E_{r}=\frac{1}{4}\sum_{\mathbf{k}\in \overline{\text{BZ}}}\mbox{tr}\,[E(\mathbf{k})-h(\mathbf{k})]
\end{equation}
is the zero-point energy correction due to quantum fluctuations. Using this result,
we obtain the quantum selection of the quadrupolar order in the F$_{\text{xy}}$
state and the 120$^{\circ}$ N\'{e}el state.

\begin{figure*}[t]
\centering
\includegraphics[width=.325\textwidth]{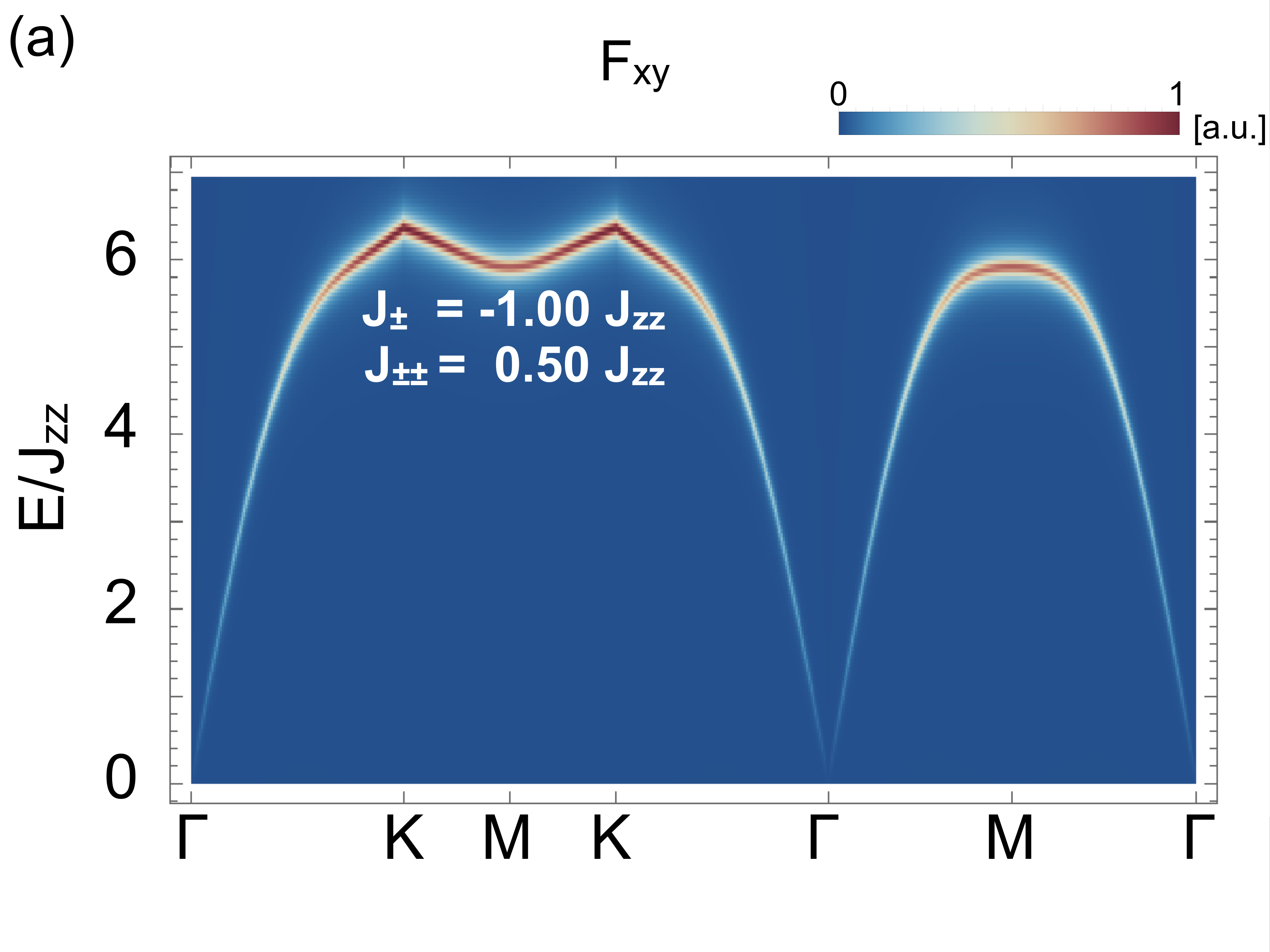}
\includegraphics[width=.325\textwidth]{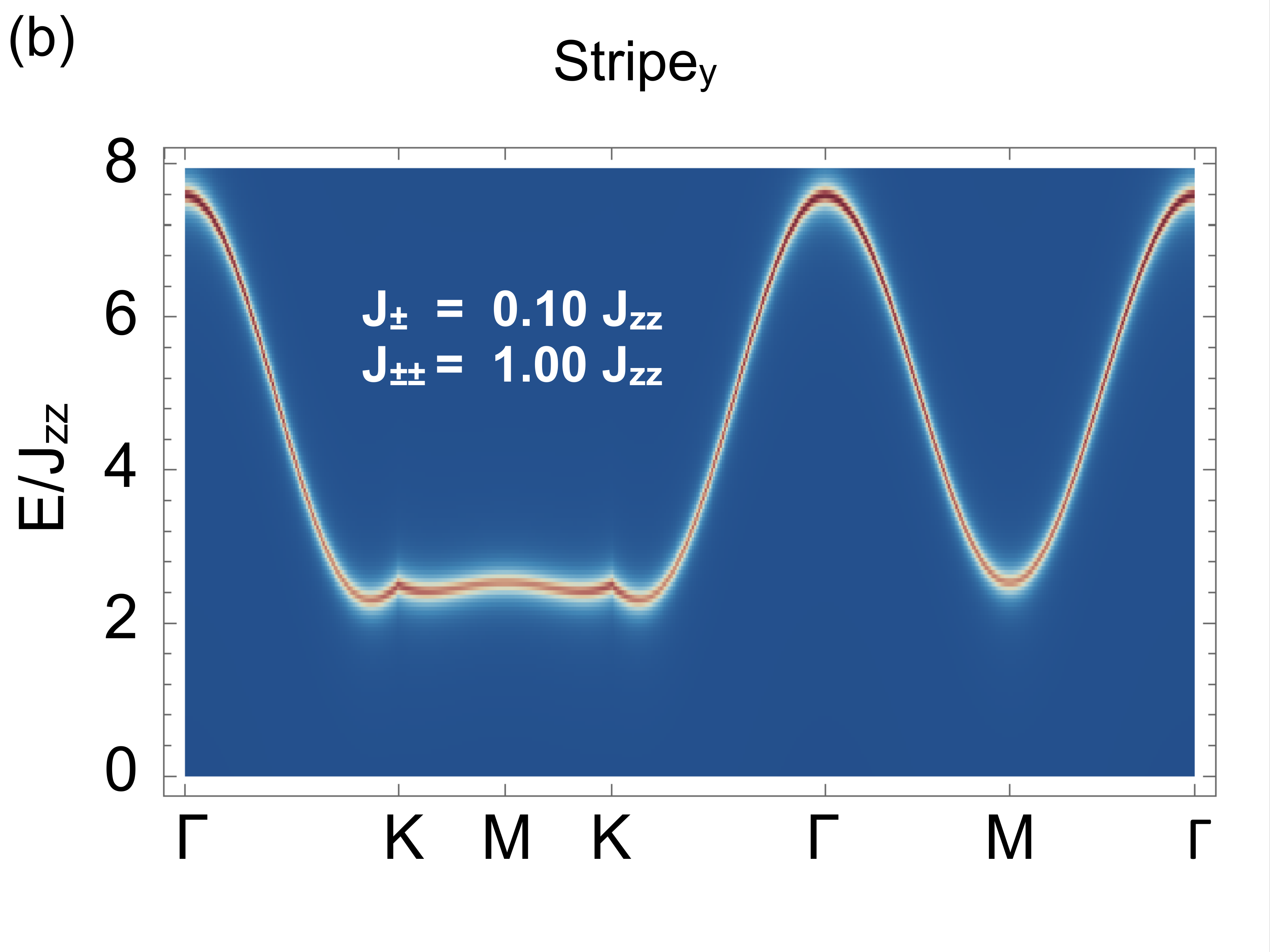}
\includegraphics[width=.325\textwidth]{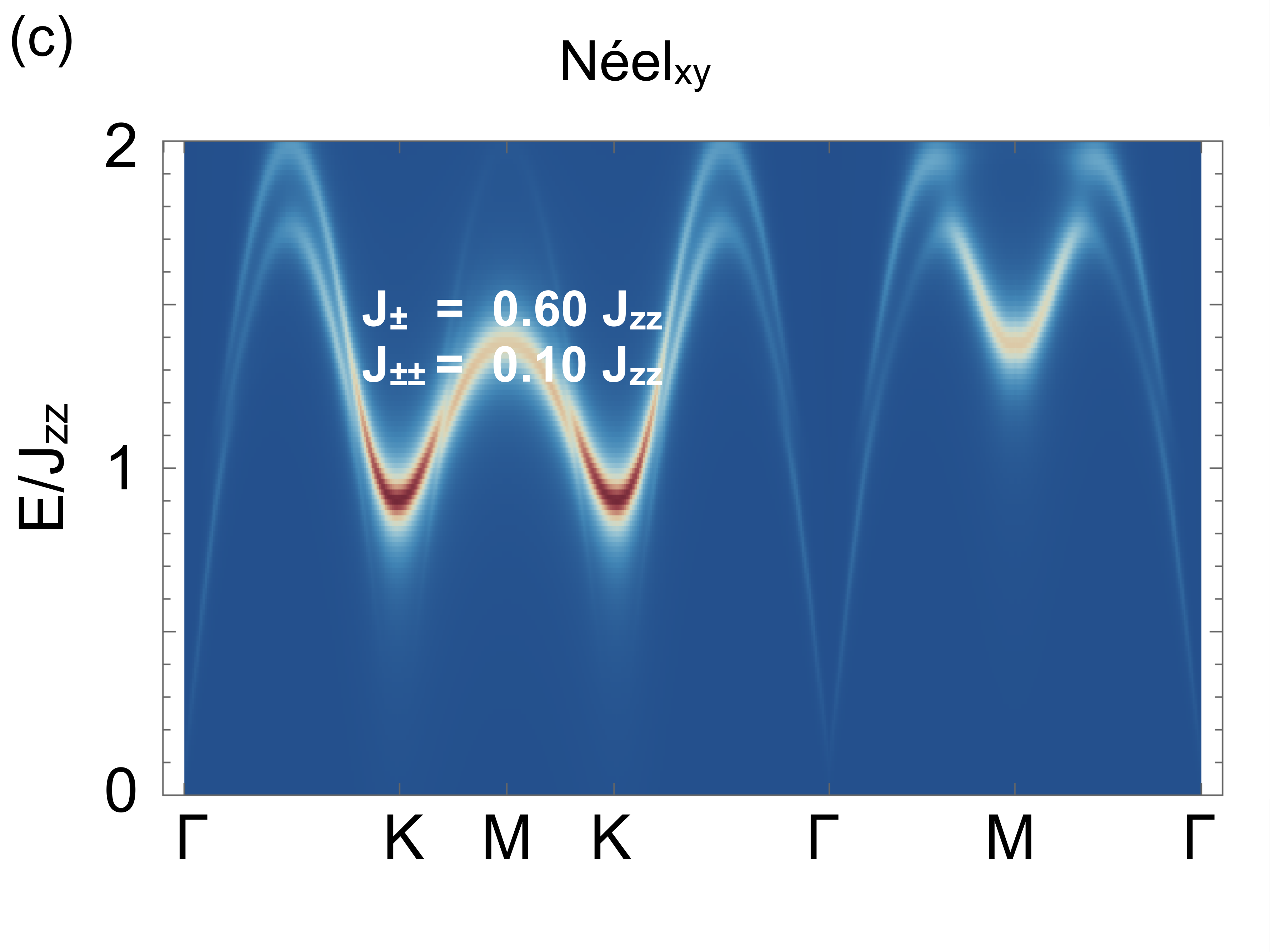}
\includegraphics[width=.325\textwidth]{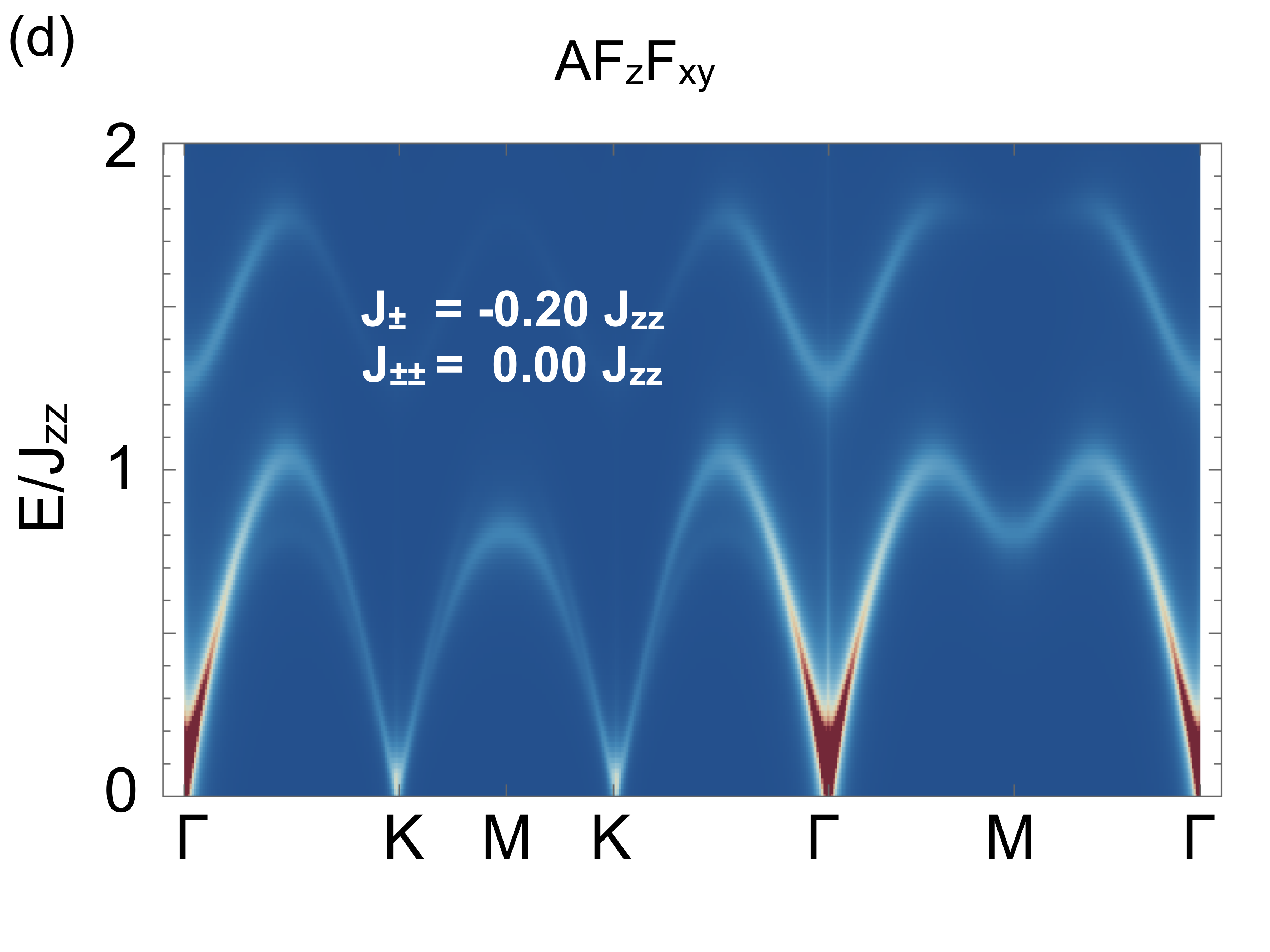}
\includegraphics[width=.325\textwidth]{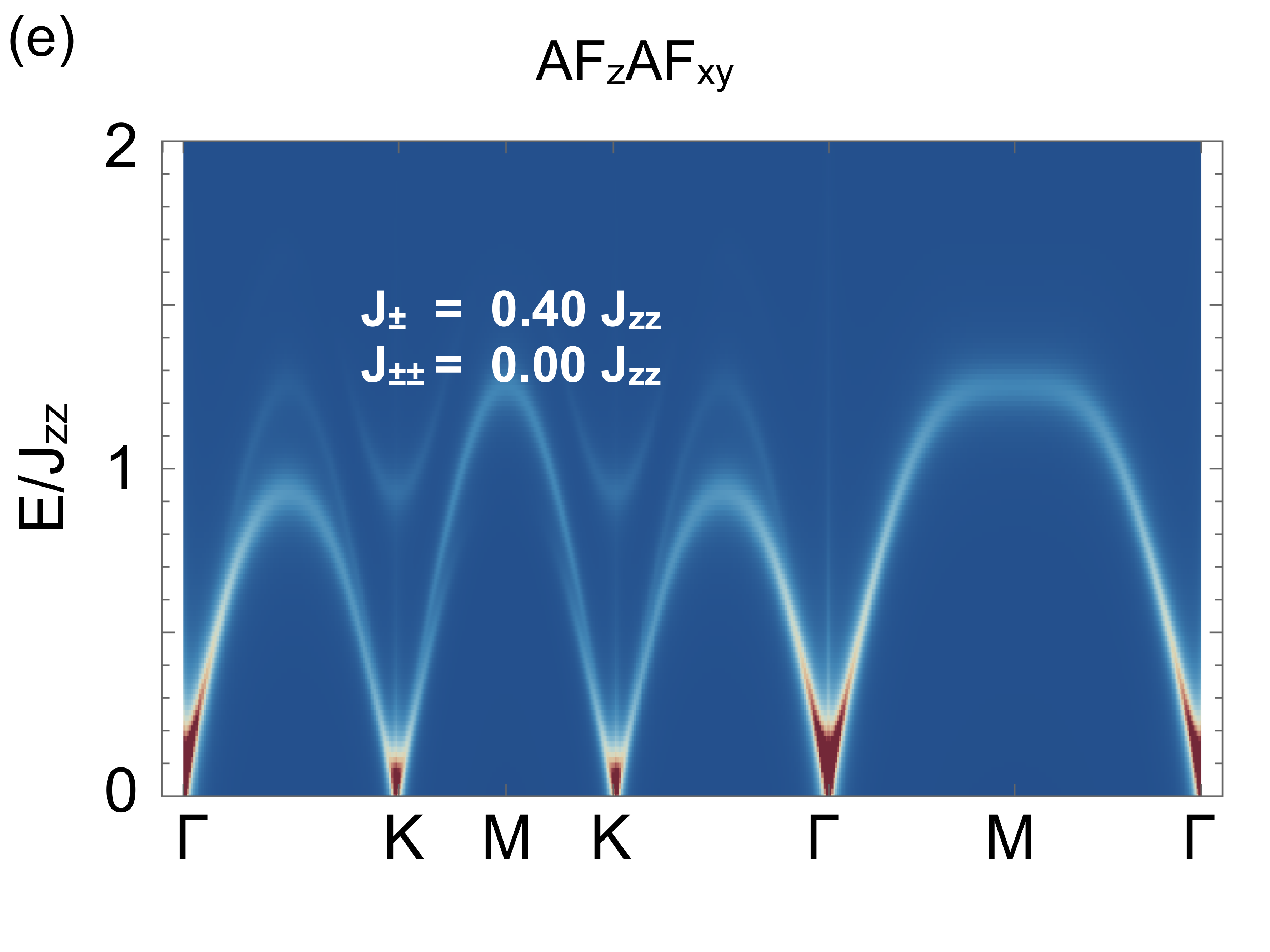}
\includegraphics[width=.325\textwidth]{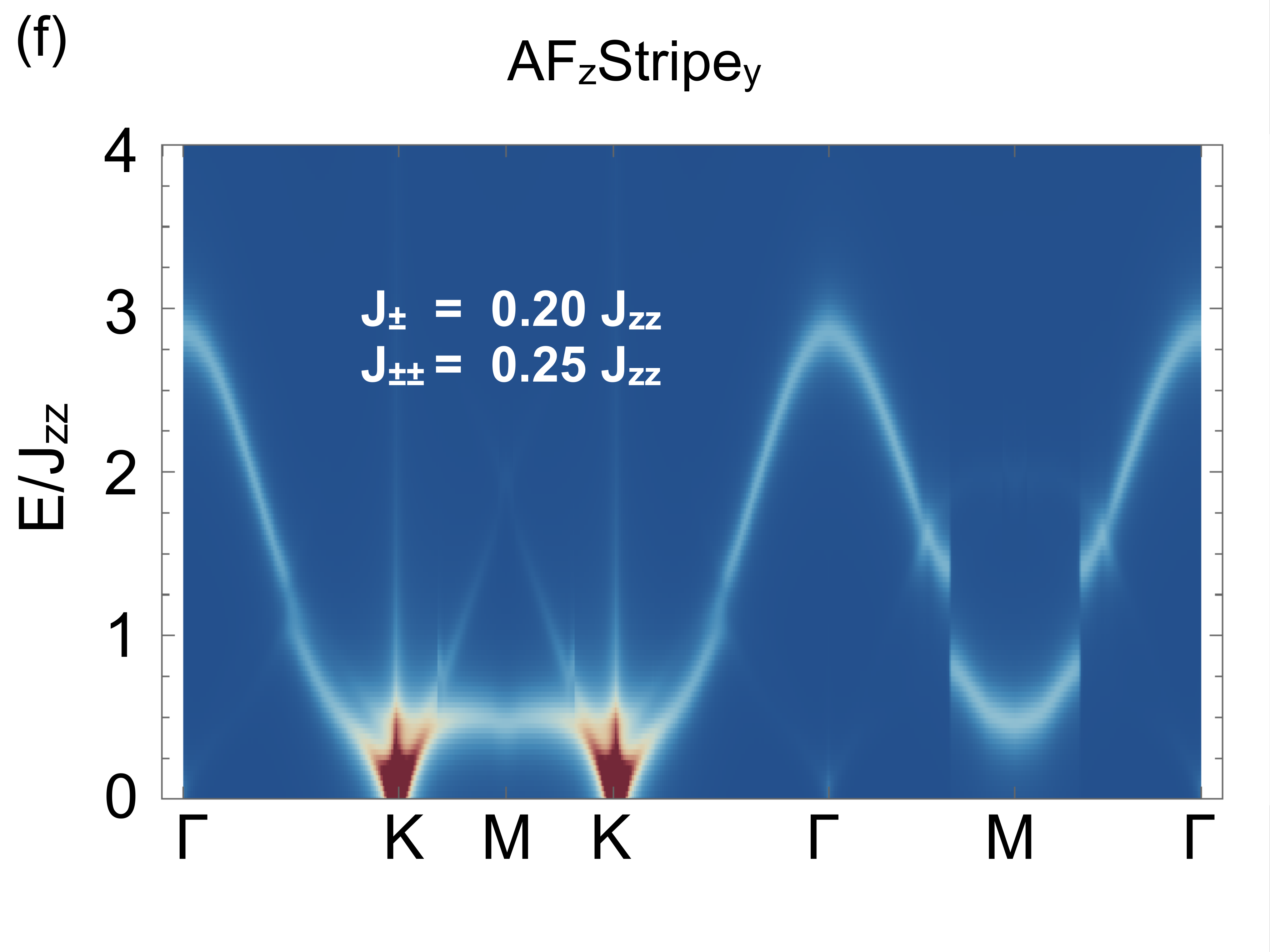}
\caption{
Dynamic spin structure factors for the phases discussed in Sec.~\ref{sec3}, obtained from the linear spin wave theory.
The representative parameters for different subfigures are given. The plots here are intensity plots. 
We also plot the full spin wave dispersions in Appendix.~\ref{appendix3}.
}
\label{fig6}
\end{figure*}

\section{Detection of multipolar orders and excitations}
\label{sec5}

As we have already indicated in the previous sections, the quadrupolar order is not
directly visible from the conventional magnetic measurement. Instead, the dynamical
measurement is able to observe the consequence of the quadrupolar orders. What is
essential here is the non-commutative relation between the dipole component and the
quadrupole component. It is this property that manifests the dynamics of the quadrupolar
order in the $S^z$ correlator. The dipole component, $S^z$, couples linearly with
the external magnetic field. Likewise, the neutron spin would only couple to the
dipole moment $S^z$ at the linear order. Therefore, the inelastic neutron scattering
would measure the $S^z$-$S^z$ correlation,
\begin{eqnarray}
\label{eq3}
&&{\mathcal S}^{zz}(\mathbf{q},\omega > 0) \nonumber \\
&& \quad  =\frac{1}{2\pi N}\sum_{ij}\int_{-\infty}^{+\infty}\mbox{d}t\,e^{i\mathbf{q}\cdot(\mathbf{r}_{i}-\mathbf{r}_{j})-i\omega t}\langle S_{i}^{z}(0) S_{j}^{z}(t)\rangle.
\end{eqnarray}
In this section, we discuss the dynamic information of the system that is encoded in the inelastic neutron scattering measurements.

The remarkable feature of the {\sl selective} coupling of the neutron spins to
the magnetic moments greatly facilitates the identification of the intertwined
multipolar orders. One can separately read off signatures of the ordering of
dipole and quadrupole moments from elastic and inelastic neutron scattering
measurements, respectively. The latter is because the $S^z$ moment creates
spin-flipping events on the quadrupole moments and thus creates coherent
spin-wave excitations. These excitations then carry the information about
the underlying quadrupolar ordering structures. Thus, although the quadrupolar
moments do not directly couple to the magnetic field, the quadrupolar excitations
can be indirectly probed. The dynamic spin structure factor, that is defined
in Eq.~\eqref{eq3} and measured by inelastic neutron scattering, encodes the
dispersion and intensity of the quadrupolar excitations. In the following we
use the linear spin wave theory to calculate the dynamic spin structure factor.
We follow Ref.~\onlinecite{toth2015} and find that at zero temperature the
dynamic spin structure factor takes the form
\begin{eqnarray}
&& {\mathcal S}^{zz}(\mathbf{k},\omega > 0)
\nonumber \\
&& \quad 
=\frac{S}{2M}\sum_{s=1}^{M}[T_{\mathbf{k}}^{\dagger}\mathbf{U}^{z}(\mathbf{U^{z})}^{\dagger}T_{\mathbf{k}}]_{s+M,s+M}\delta(\omega-\omega_{\mathbf{k}s}),
\end{eqnarray}
where the $2M$-dimensional vector $\mathbf{U}^{z}$ is defined as
\begin{eqnarray}
\mathbf{U}^{z}&=&[u_{1}^{z}+iv_{1}^{z},u_{2}^{z}+iv_{2}^{z},\cdots,u_{M}^{z}+iv_{M}^{z},
\nonumber
\\ && \,\, u_{1}^{z}-iv_{1}^{z},u_{2}^{z}-iv_{2}^{z},\cdots,u_{M}^{z}-iv_{M}^{z}]^{T}.
\end{eqnarray}
Here, we have neglected the two-magnon process in the above expression.

Our results are displayed in Fig.~\ref{fig6}. The gapless modes in the figures are pseudo-Goldstone modes that arise from the emergent continuous degeneracy at the mean-field level and the linear spin-wave treatment. High order quantum fluctuations would create a mini-gap for these modes. Despite that, we expect a $T^2$ heat capacity behavior for the temperature regime above the mini-gap energy scale in the ordered phase.

For our model that describes the spin-1/2 degrees of freedom, the number of magnon branches
should be equal to the number of sublattices in the corresponding ordered phase. However, surprisingly we find
that for two-sublattice Stripe$_{\text y}$ and six-sublattice AF$_{\text z}$Stripe$_{\text y}$
structures, we can see only see one and three bands respectively,
which implies that half of the bands are completely invisible in the
$S^{z}$-$S^{z}$ correlator (see Appendix.~\ref{appendix3} for a comparison).
The underlying reason is the selection
rule associated with the symmetry generated by
\begin{equation}
\hat{W} = T_{-{\bf a}_1 + {\bf a}_2}\otimes e^{i \pi \sum_j S_j^z}.
\end{equation}
where 
$T_{- {\bf a}_1 + {\bf a}_2}$ denotes the lattice translation by ${- {\bf a}_1 + {\bf a}_2}$.
The Hamiltonian stays invariant under $\hat{W}$, $[\hat{W}, H ] = 0$.

From now on, we introduce the notation $s$ and $\bar{s}$ to denote the sublattice pair that are interchanged under the action of $\hat{W}$. In the labelling of Fig.~\ref{fig4}, we find that $\bar{A} = B, \bar{C} = D, \bar{E} = F$.

For the elementary excitations, the effect of $\hat{W}$ is such that
\begin{eqnarray}
\text{Stripe$_{\text y}$}: \hat{W} b_{{\bf k}, s} \hat{W}^\dagger  &=& e^{i\phi({\bf k})} b_{{\bf k}, \bar{s}}, s = A, B \quad \\
\text{Stripe$_{\text y}$AF$_{\text z}$}: \hat{W} b_{{\bf k}, s} \hat{W}^\dagger  &=& e^{i\phi({\bf k})} b_{{\bf k}, \bar{s}}, s = A, \ldots, F,\quad
\end{eqnarray}
where $\phi({\bf k}) = - k_x + k_y$.

The eigenmodes of $\hat{W}$ take bonding/antibonding form,
\begin{eqnarray}
    \alpha_{{\bf k}, s, \pm} = b_{{\bf k}, s} \pm b_{{\bf k}, \bar{s}},
\end{eqnarray}
whose eigenvalues are
\begin{eqnarray}
    \hat{W} \alpha_{{\bf k},s,\pm} \hat{W}^\dagger = \pm e^{i\phi({\bf k})} \alpha_{{\bf k},s,\pm}.
\end{eqnarray}
Since $\hat{W}$ is a symmetry of the Hamiltonian, the energy eigenmodes are separate linear combinations of $\alpha_{{\bf k}, s, \pm}$,
\begin{eqnarray}
    \beta_{{\bf k}, t, \pm} = \sum_{s} c_{t,s} \alpha_{{\bf k}, s, \pm} + d_{t,s} \alpha^\dagger_{-{\bf k}, s, \pm}
\end{eqnarray}
and
\begin{eqnarray}
    \hat{W} \beta_{{\bf k},t,\pm} \hat{W}^\dagger = \pm e^{i\phi({\bf k})} \beta_{{\bf k},t,\pm}.
\end{eqnarray}
The $\pm$ branches do not mix, since they have distinct eigenvalues under $\hat{W}$.

On the other hand, we can make a spectral representation of Eq.~(\ref{eq3}) as follows
\begin{widetext}
\begin{eqnarray}
S^{zz}({\bf q}, \omega > 0) &=& \sum_n \left\langle 0 \left| \sum_{s=1}^M S_s^z(-{\bf q}, -\omega) \right| n \right\rangle
\left\langle n \left| \sum_{s=1}^M S_s^z({\bf q}, \omega) \right| 0 \right\rangle \nonumber \\
&\propto& \sum_n \delta\left(\omega-(\epsilon_n-\epsilon_0)\right) \left\langle 0 \left| \sum_{s=1}^M (b_{{\bf q}, s} + b_{-{\bf q}, s}^\dagger) \right| n \right\rangle
\left\langle n \left| \sum_{s=1}^M (b_{-{\bf q}, s} + b^\dagger_{{\bf q}, s}) \right| 0 \right\rangle \nonumber\\
&\propto& \sum_n \delta\left(\omega-(\epsilon_n-\epsilon_0)\right) \left\langle 0 \left| \sum_{s=1}^M (\alpha_{{\bf q}, s, +} + \alpha^\dagger_{-{\bf q}, s, +}) \right| n \right\rangle
\left\langle n \left| \sum_{s=1}^M (\alpha_{-{\bf q}, s, +} + \alpha^\dagger_{{\bf q}, s, +}) \right| 0 \right\rangle.
\end{eqnarray}
\end{widetext}
It is thus obvious that the contribution is nonzero if and only if $\ket{n}$ is created by the $\beta_{{\bf k}, t, +}$ operators.
The $\beta_{{\bf k}, t, -}$ states are not accessible.
As a result, the $S^{z}$-$S^{z}$ correlation
function only measures coherent excitations with even parity. The
odd parity excitations, instead, are present in $S^{x}$-$S^{x}$
and $S^{y}$-$S^{y}$ correlation functions.

The elastic neutron scattering measurement directly probes the magnetic ground state of the $S^z$ components. The ordering wave vector of the dipolar moment $S^z$ will be the magnetic Bragg peak in the static spin structure factor. For states with pure quadrupolar orders like F$_{\text{xy}}$, Stripe$_{\text y}$, and N\'{e}el$_{\text{xy}}$, there is no dipolar ordering and the ground state does not break time reversal symmetry, so there are no Bragg peaks in static spin structure factors. For states with intertwined multipolar orders such as AF$_{\text z}$F$_{\text{xy}}$, AF$_{\text z}$AF$_{\text{xy}}$, and AF$_{\text z}$Stripe$_{\text y}$, however, the dipolar components order into a multi-sublattice pattern. The unit cell for the dipolar order is effectively enlarged, and hence one should observe the magnetic Bragg peaks at the $K$ point in the Brillouin zone.

\section{The magnetization process}
\label{sec6}

The peculiar property of the quadrupolar order and the non-Kramers doublets also
lies in the magnetization process of the system under the external magnetic field.
Although the magnetic field does not directly couple to the quadrupole components
of the local moment, the magnetization is influenced by the underlying quadrupolar
order. The behavior of the magnetization should provide information about the
hidden quadrupolar order that is otherwise not directly measurable. To explore
this idea, we first introduce the magnetic field to the system so that we have
\begin{eqnarray}
H_{h} & = & H-g\mu_{\text{B}}h\sum_{i}S_{i}^{z}
\nonumber \\
& \equiv & H-B\sum_{i}S_{i}^{z}.
\end{eqnarray}
From the expression of the above Hamiltonian, one can immediately read off the
Curie-Weiss temperature. Because the external magnetic field only couples to
the $S^{z}$ component of the local moment, the Curie-Weiss temperature only
reflects the $J_{zz}$ interaction, i.e.
\begin{eqnarray}
\Theta_{\text{CW}}=-\frac{3}{2}J_{zz}.
\end{eqnarray}

\begin{figure*}[t]
\centering
\includegraphics[width=.24\textwidth]{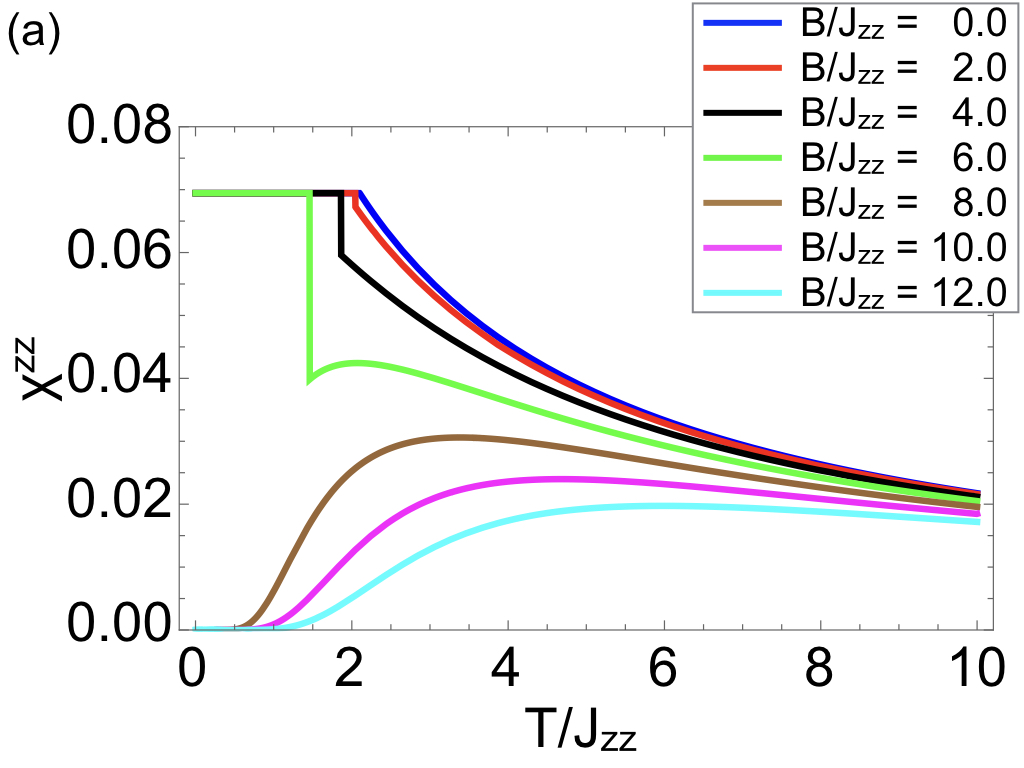}
\includegraphics[width=.24\textwidth]{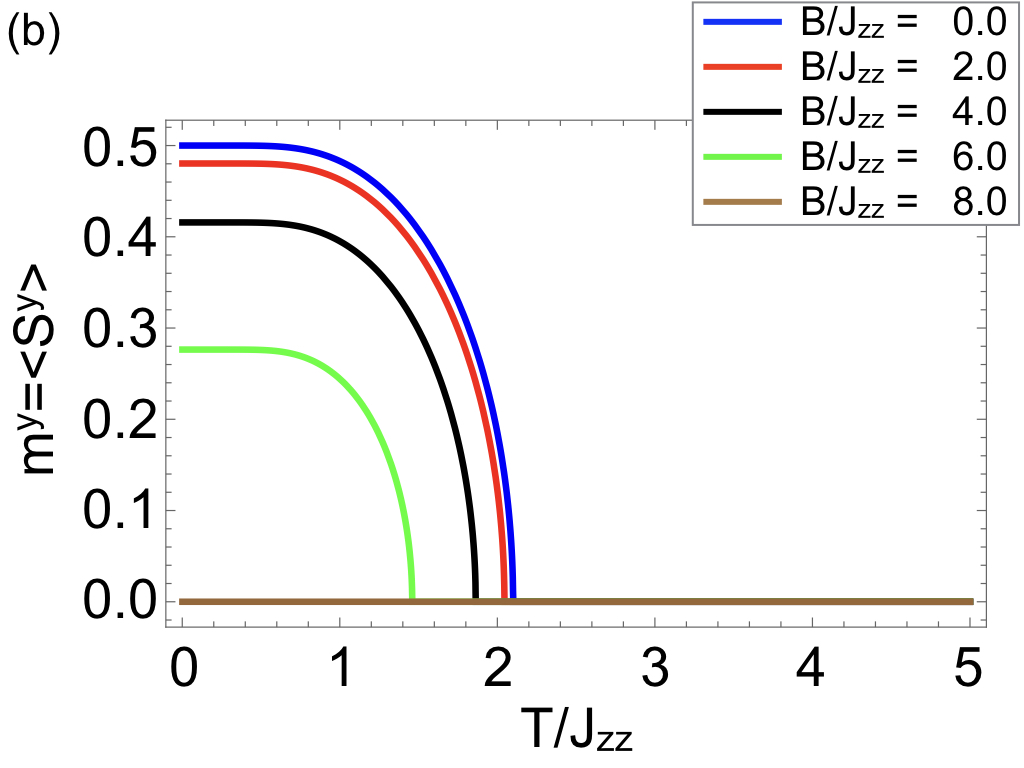}
\includegraphics[width=.24\textwidth]{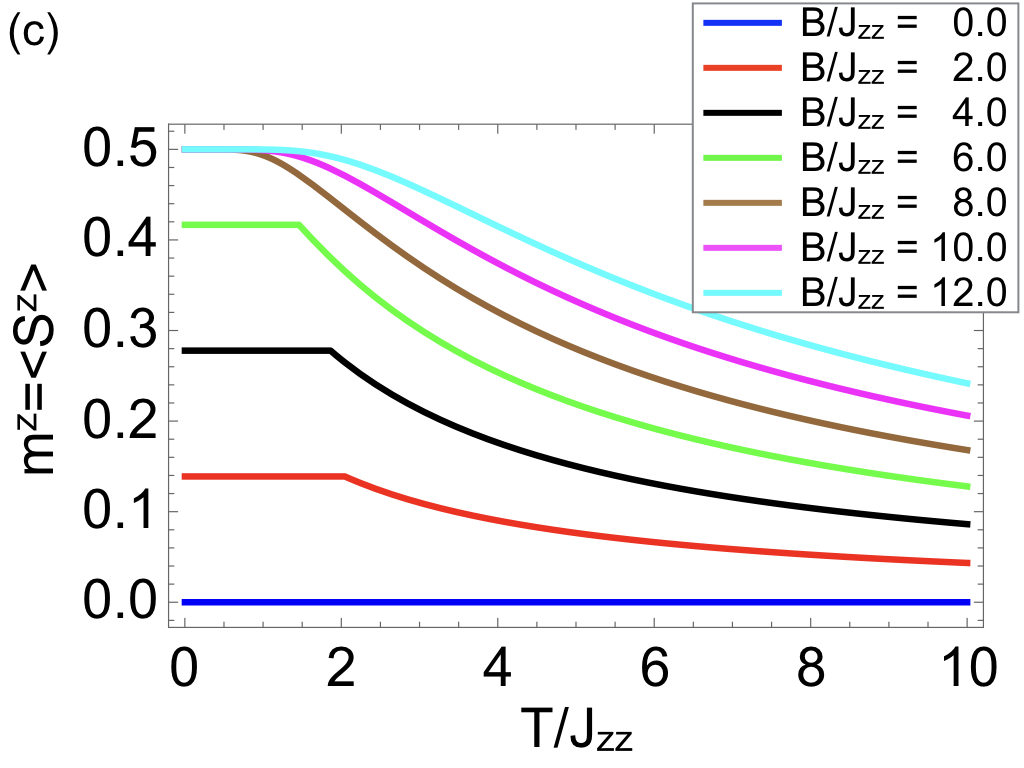}
\includegraphics[width=.24\textwidth]{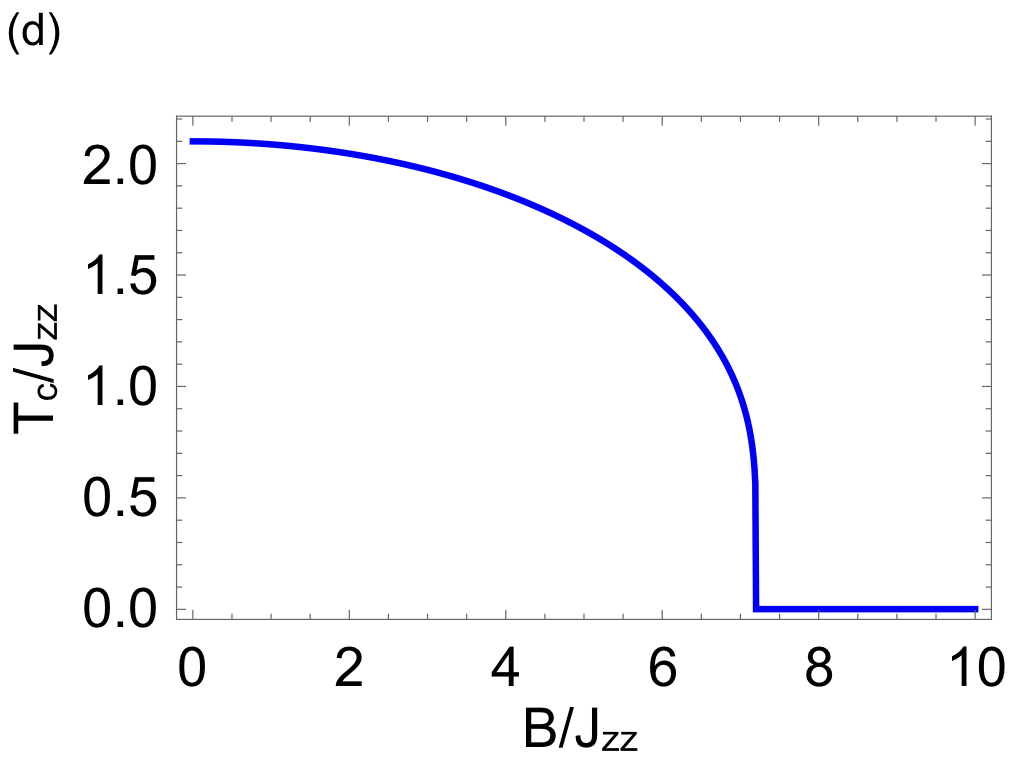}
\caption{
(a) The magnetic suscpetibility $\chi^{zz}$ at finite temperatures for different 
field strengths. There is a transition at ${B_c \simeq 7.2 J_{zz}}$. (b) The 
antiferro-quadrupolar order parameter at finite transversal fields and 
finite temperatures. (c) The dipolar moment at finite fields and finite temperatures.
(d) The ordering temperature of the antiferro-quadrupolar Stripe$_{\text y}$ order 
as a function of transversal fields. In all figures we choose ${J_{\pm} = 0.1J_{zz}}$, and ${J_{\pm\pm} = J_{zz}}$, so that the ground state at zero field is in the Stripe$_{\text y}$ phase.
}
\label{fig7}
\end{figure*}

The impact of the underlying quadrupolar order on the magnetization 
should be most clear for the pure quadrupolar ordered phase. Here,
we explore the physics on the Stripe$_{\text y}$ state. The field 
will polarize the dipolar moments and suppress the quadrupolar ordering. 
In Fig.~\ref{fig7}, we choose the coupling constants deep in the 
antiferro-quadrupolar Stripe$_{\text y}$ phase, where 
${J_{\pm} = 0.1 J_{zz}}$ and ${J_{\pm\pm} = 1.0 J_{zz}}$. 
The mean-field ansatz is chosen to take care of the uniform 
$S^z$ magnetization,
\begin{eqnarray}
\left\langle \bs{S}_i \right\rangle \equiv
\left[
m^x_i, m^y_i, m^z_i
\right]^T =
\left[
0, e^{i \bs{M} \cdot \bs{R}_i} m^y, m^z
\right]^T,
\end{eqnarray}
where ${\bs{M}=\left(0, {2\pi}/{\sqrt{3}}\right)}$ is the ordering 
wave vector for the Stripe$_{\rm y}$ state, and $m^y$, $m^z$ real 
numbers subject to the constraint ${\left| \left\langle \bs{S}_i
\right\rangle\right| = S}$ for all sites $i$.

From the mean-field analysis, we plot the magnetization at zero and 
finite temperatures for different strengths of transverse fields. 
In addition, the magnetic susceptibility and the dependence of the 
ordering temperature on the strength of the magnetic field are shown 
together in Fig.~\ref{fig7}. We first discuss the zero field 
susceptibility $\chi^{zz}$ (see Fig.~\ref{fig7}a). Because of the 
lack of dipolar ordering, there is a constant $\chi^{zz}$ below $T_c$, 
and smoothly decays below $T_c$, obeying the Curie-Weiss law. In particular, 
it does not develop a peak across the finite-temperature transition at $T_c$, 
because the quadrupolar order paramter is ``hidden'' to the magnetic field. 
This is to be contrasted with the case of Kramers doublets, where the 
susceptibily shows a critical behavior at $T_c$.

The magnetic field suppresses the antiferro-quadrupolar order. This is because the magnetization does not commute with the quadrupolar order parameter $S^y$. When the field polarizes the magnetization, the quantum fluctuation of the quadrupolar orders is enhanced and thereby reducing the ordering temperature. This physics has also been suggested for the electronic multipolar orders in intermetallic compounds TmAu$_2$ and TmAg$_2$, where the lattice strain is introduced to control the electronic quadrupolar order~\cite{Kivelson}. Here we introduce the magnetic field to control the magnetization. In Fig.~\ref{fig7}b and Fig.~\ref{fig7}c, we explicitly show this result from our mean-field theory.

Above the critical field $B_c$, the quadrupolar order is completely suppressed, and the dipolar moments are polarized by the field. The $\mathbb{Z}_2$ symmetry is generated by the rotation operation
\begin{eqnarray}
    \hat{U} \equiv e^{i\pi S^z}.
\end{eqnarray}
It is spontaneously broken in the antiferroquadrupolar ground states by $\langle{S^y}\rangle$ and is restored in the fully magnetized state. Around $B_c$, we expect a quantum critical region and quantum phase transition due to the breaking of the $\mathbb{Z}_2$ symmetry. Although the order parameter $S^y$ cannot be directly measured, the quantum phase transition can be observed from the magnetic susceptibility at the finite $B$. In Fig.~\ref{fig7}d, we plot 
the transition temperature as a function of the external magnetic field.

The above analysis can be readily extended to F$_{\text{xy}}$, N\'{e}el, and other intertwined multipolar ordered phases.
We have chosen the Stripe$_{\text y}$ order as a representative.

\section{Discussion}
\label{sec7}

In this paper, we have established the phase diagram of the generic
interacting model that is relevant for non-Kramers doublets on the
triangular lattice. We find the broad existence of the pure quadrupolar
orders and the intertwined multipolar orders in the phase diagram.
Although the quadrupolar order is invisible from the conventional
magnetic measurements, its presence shows up in the dynamic property
and the magnetic excitations of the system. In contrast, the
dipolar order can be directly measured by conventional magnetic probes.
These unusual properties arise naturally from the selective coupling or
measurement and the non-commutative relation between the dipolar and the
quadrupolar moments. One could ``read off'' the dipolar order from the
static magnetic probes and the quadrupolar order from the dynamic magnetic
probes such as inelastic neutron scattering measurement.

In addition, we uncover the physics of the quantum order by disorder
and the related pseudo-Goldstone modes. The consequences of this result
appear both in the thermodynamics and the dynamic properties.
Apart from the nearly gapless pseudo-Goldstone modes that show up in the
dynamic measurements, the system in the relevant phases would have
a $T^2$ heat capacity.

Here we point out the material's relevance of our theoretical results.
As we have mentioned in the beginning and in Sec.~\ref{sec2}, there exists an
abundance of the triangular lattice rare-earth magnets. Most of these compounds
have not been studied carefully, maybe not even been studied in any previous work.
Among these materials, only the spin liquid candidate material YbMgGaO$_4$ has
been studied extensively, both experimentally and theoretically. For others
such as RCd$_{3}$P$_{3}$, RZn$_{3}$P$_{3}$, RCd$_{3}$As$_{3}$, RZn$_{3}$As$_{3}$,
KBaR(BO$_{3}$)$_{2}$, and many ternary chalcogenides (LiRSe$_2$, NaRS$_2$,
NaRSe$_2$, KRS$_2$, KRSe$_2$, KRTe$_2$, RbRS$_2$, RbRSe$_2$, RbRTe$_2$,
CsRS$_2$, CsRSe$_2$, CsRTe$_2$, etc)~\cite{triangle1, triangle2, triangle3,triangle4,OHTANI,Sato}, 
little is known. When R$^{3+}$ supports
a non-Kramers doublet, our model and results can be directly applied.
This could occur if R$^{3+}$ ion is the Pr$^{3+}$, Tm$^{3+}$, or Tb$^{3+}$
ion where the rare-earth ion contains even number of 4f electrons per site.
The single crystal TmMgGaO$_4$ was recently synthesized~\cite{cavaTMGO,1804.00696}, 
and the lowest energy states of the Tm$^{3+}$ ion were either a non-Kramers doublet 
or two nearly degenerate singlets\cite{1804.00696}. 
In the latter case, the effective spin model should take into account
of the effect from the crystal field splitting between two nearly degenerate 
singlets. Thus, the resulting model~\cite{Changleunpub} would be fundamentally different 
from the model in Eq.~\eqref{eq1}. To resolve the single-ion ground state 
as well as the many-body ground state for TmMgGaO$_4$, further crystal electric 
field study and experimental efforts are required.

Finally, our proposal for the selective measurement of the intertwined multipolar
orders is not just specific to the non-Kramers doublets on triangular lattice.
This piece of physics could be well extended to the non-Kramers doublets on
other lattices. More broadly, any physical system with intertwined multipolar
orders or hidden orders could potentially hold this kind of physics and phenomenon.
The idea of {\sl using non-commutative observables} to probe the dynamics of the
hidden orders is central to our proposal and should be well extended to many other
systems. These not only include the magnetic multipolar orders that are discussed in
this paper but also contain the electronic analogues of the multipolar orders
that have been discussed for example for the tetragonal intermetallic compounds
TmAu$_2$ and TmAg$_2$ in Ref.~\onlinecite{Kivelson} and
for URu$_2$Si$_2$~\cite{RevModPhys.83.1301}.

\section{Acknowledgments}

We acknowledge an ongoing collaboration with Professor Wenan Guo's group from
Beijing Normal University, and Professor Sasha Chernyshev from University of
California Irvine for an email correspondence. This work is supported by the
ministry of science and technology of China with the Grant No.2016YFA0301001,
the start-up fund and the first-class university construction fund of Fudan
University, and the thousand-youth-talent program of China.

\appendix

\section{The relevance of our model to Kitaev interaction}
\label{appendix1}

In a previous work~\cite{Feiye_PRB2017}, we have pointed out that the vast
numbers of rare-earth magnets can support the Kitaev interaction. This means 
the Kitaev interaction goes much beyond the iridate system that was previously
proposed~\cite{Khaliullin}, and we
illustrated the observation from the rare-earth double perovskites~\cite{Feiye_PRB2017}.
Similar suggestion was made for Co-based magnets~\cite{Khaliullin2,Sano2}.
Here we extend our idea to our model on the triangular lattice.

\begin{figure}[b]
    \centering
   \includegraphics[width=3.6cm]{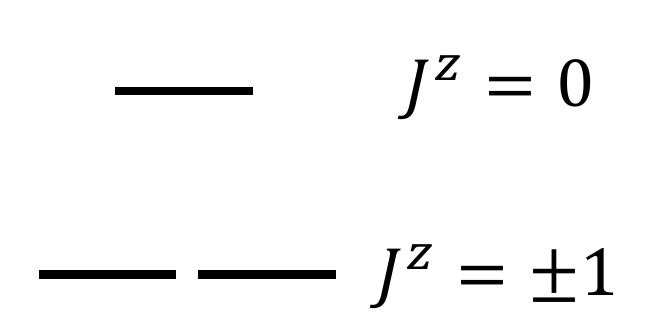}
    \caption{The non-Kramers ground state doublet from splitting
    the spin-1 triplets.}
    \label{sfig1}
\end{figure}

\begin{figure*}[t]
\centering
\includegraphics[width=.325\textwidth]{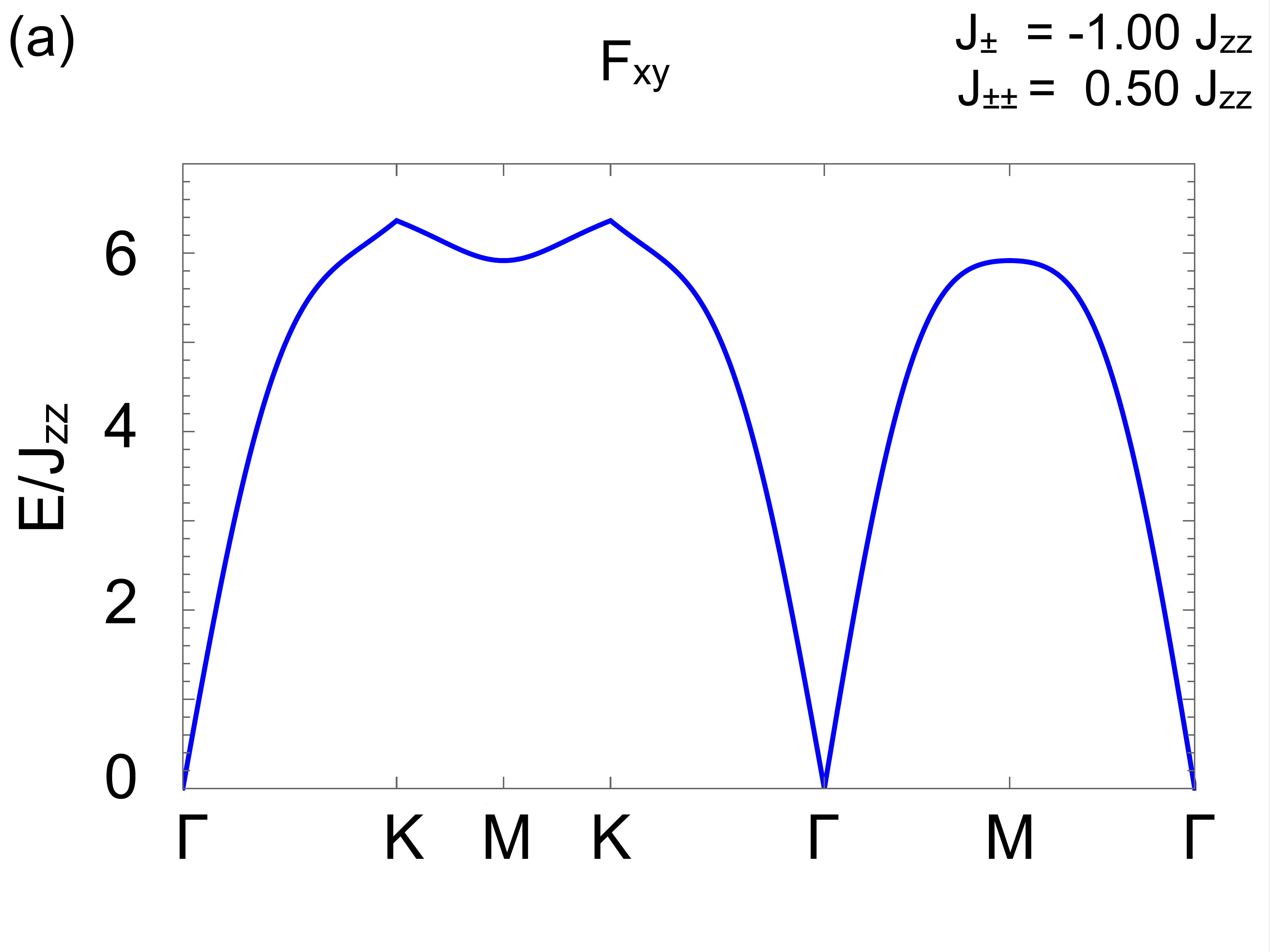}
\includegraphics[width=.325\textwidth]{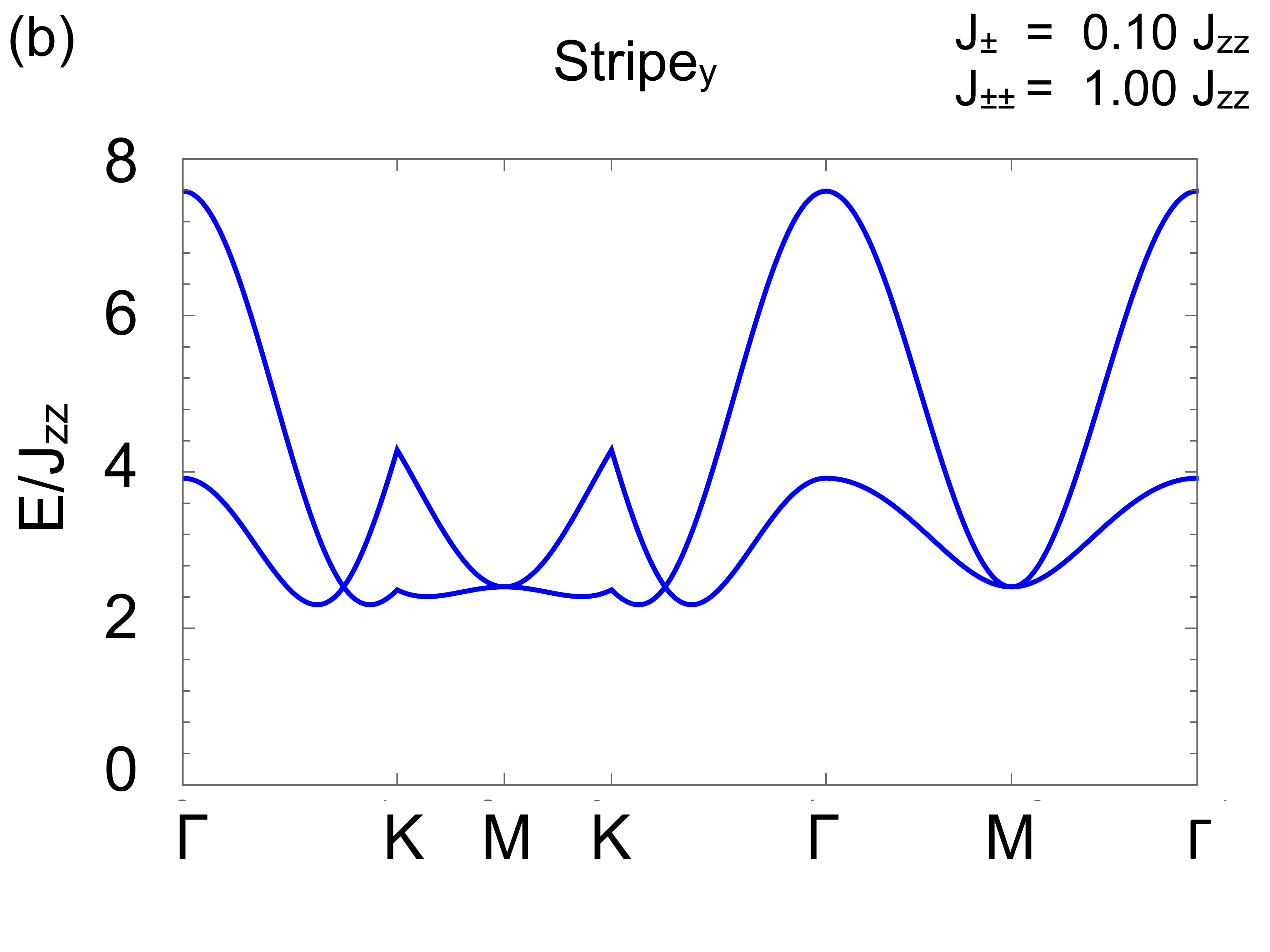}
\includegraphics[width=.325\textwidth]{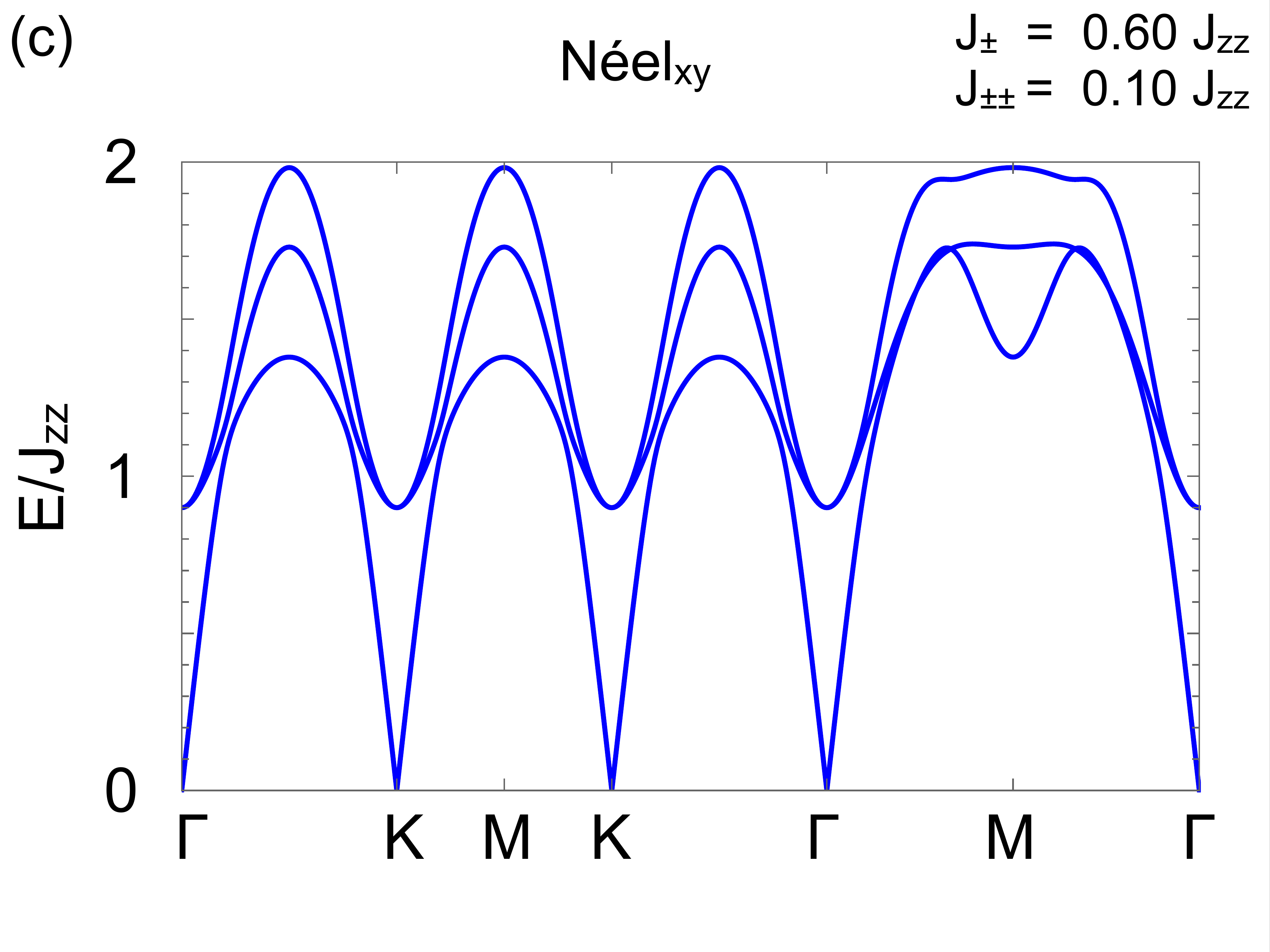}
\includegraphics[width=.325\textwidth]{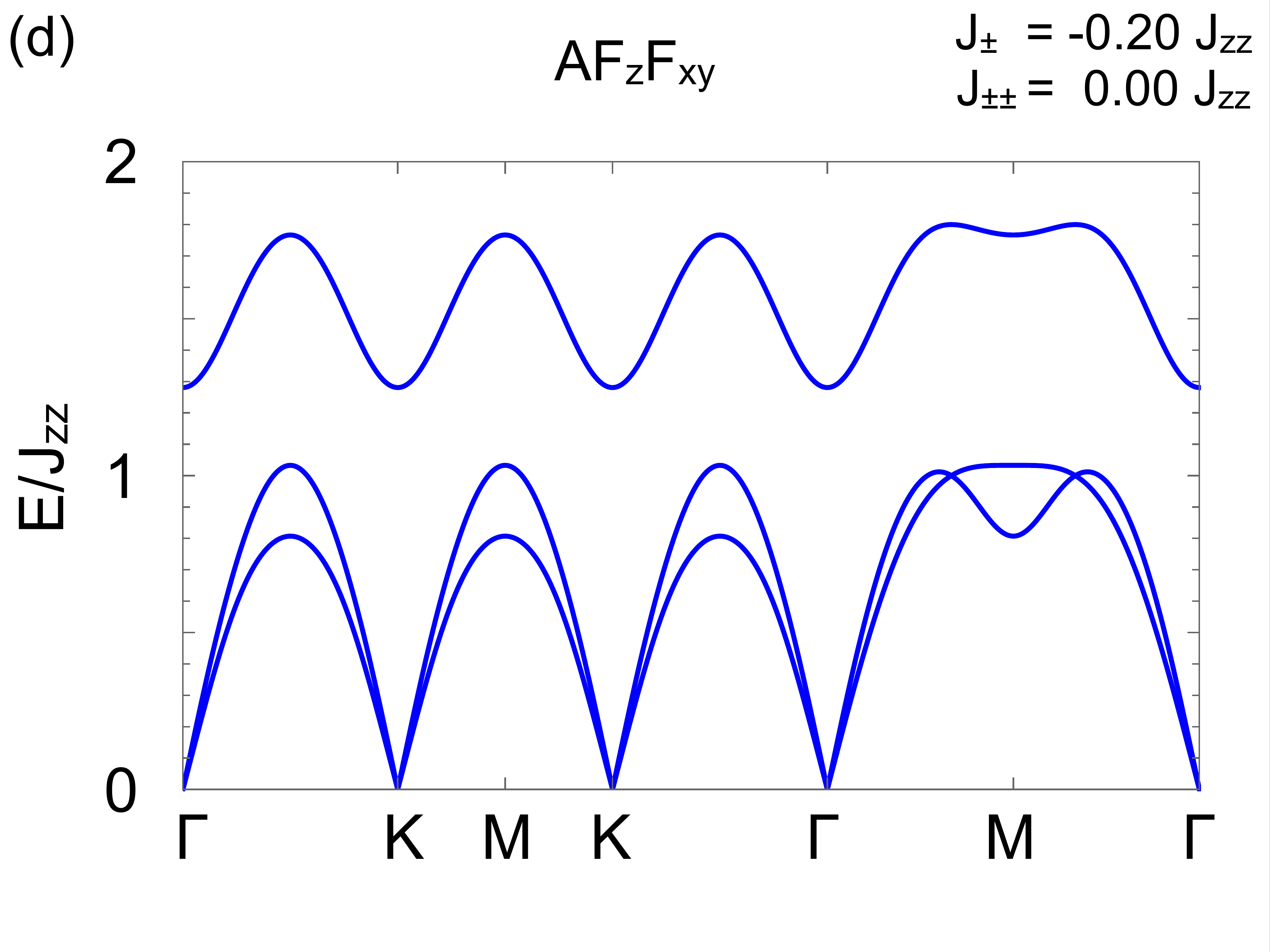}
\includegraphics[width=.325\textwidth]{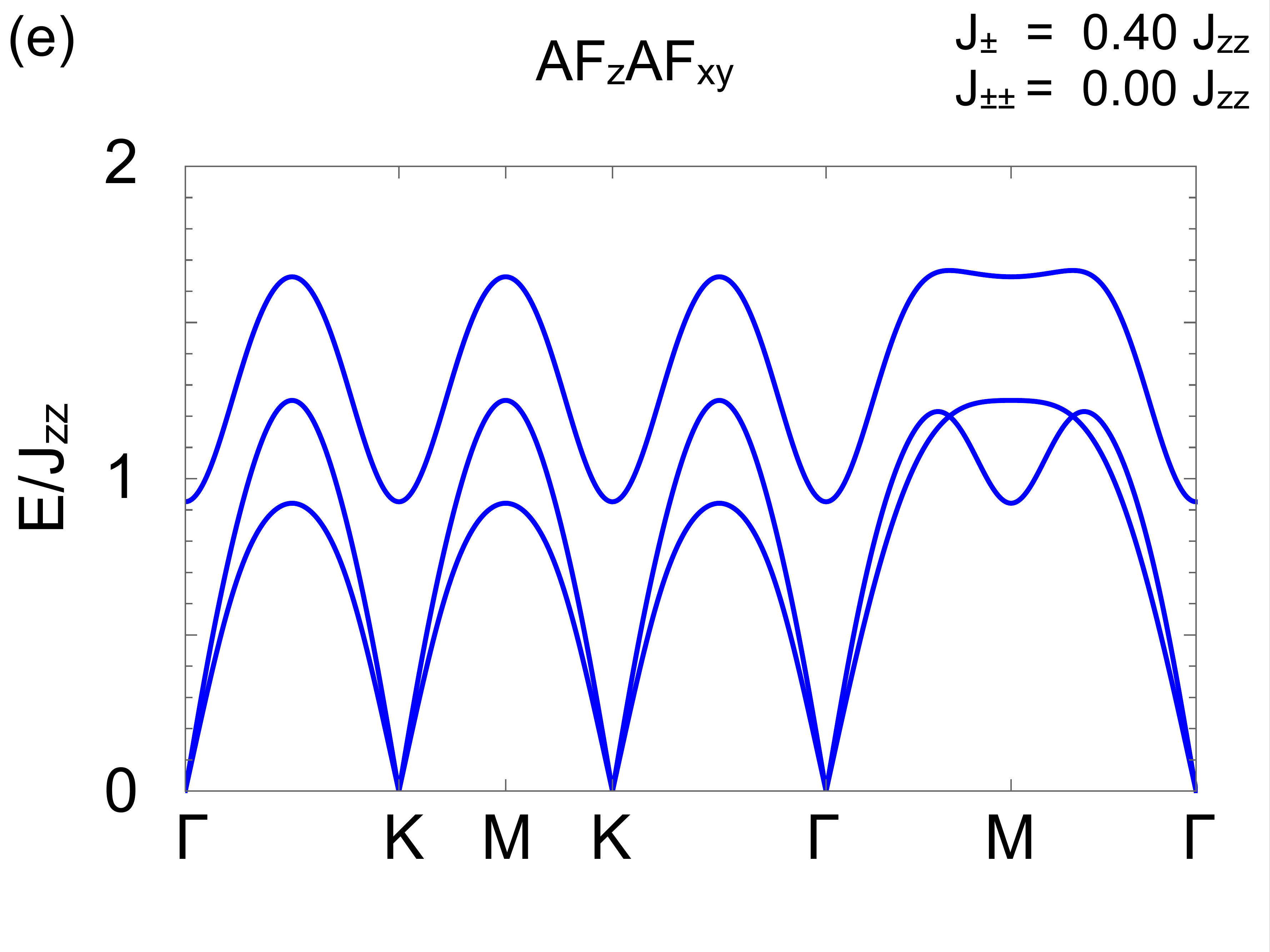}
\includegraphics[width=.325\textwidth]{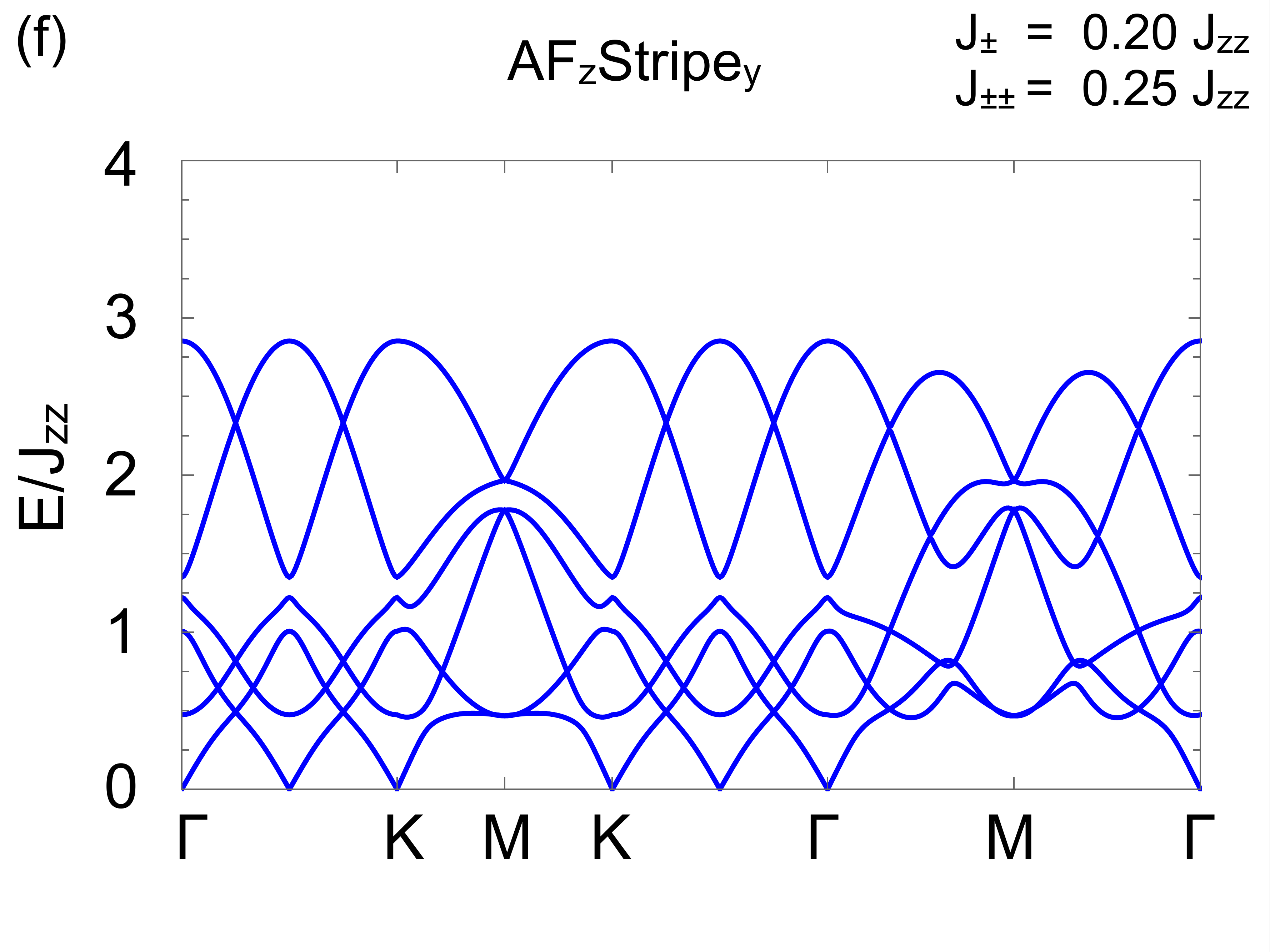}
\caption{The complete
spin-wave dispersions for different phases.
 The parameters are chosen to be the same as in Fig.~\ref{fig6}.
 The plots of the full dispersions here are to be compared 
 with the intensity plots in Fig.~\ref{fig6}.
}
\label{sfig2}
\end{figure*}

Generally speaking, a system with a local on-site three-fold rotational
symmetry and spin-orbit-entangled doublets (except dipole-octupole doublets)
would necessarily have a Kitaev interaction. These lattices include
honeycomb lattice, triangular lattice, pyrochlore lattice, FCC lattice,
{\sl et al}. Our results can be well extended to these lattices. 
The three-fold rotation permutes the bonds connecting to
the lattice site of the rotational center, and at the same time,
permutes the spin components. For our triangular lattice, we define
\begin{eqnarray}
S_{i}^{a} & \equiv & \sqrt{\frac{2}{3}}S_{i}^{x}+\sqrt{\frac{1}{3}}S_{i}^{z},\\
S_{i}^{b} & \equiv & \sqrt{\frac{2}{3}}(-\frac{1}{2}S_{i}^{x}+\frac{\sqrt{3}}{2}S_{i}^{y})+\sqrt{\frac{1}{3}}S_{i}^{z},\\
S_{i}^{c} & \equiv & \sqrt{\frac{2}{3}}(-\frac{1}{2}S_{i}^{x}-\frac{\sqrt{3}}{2}S_{i}^{y})+\sqrt{\frac{1}{3}}S_{i}^{z},
\end{eqnarray}
then our Hamiltonian in Eq.~\eqref{eq1} is recasted into the following form,
\begin{eqnarray}
H & = & \sum_{\langle ij\rangle\in\alpha}\Big[J{\boldsymbol{S}}_{i}\cdot{\boldsymbol{S}}_{j}+KS_{i}^{\alpha}S_{j}^{\alpha}\nonumber \\
 &  & +\sum_{\beta,\gamma \neq \alpha}\Gamma(S_{i}^{\alpha}S_{j}^{\beta}+S_{i}^{\beta}S_{j}^{\alpha}+S_{i}^{\alpha}S_{j}^{\gamma}+S_{i}^{\gamma}S_{j}^{\alpha})\nonumber \\
 &  & +\sum_{\beta,\gamma \neq \alpha}(K+\Gamma)(S_{i}^{\beta}S_{j}^{\gamma}+S_{i}^{\gamma}S_{j}^{\beta})
 \Big],
\end{eqnarray}
where $\alpha= a, b, c$ labels the nearest-neighbor bond, the $a$ bond
corresponds to the ${\boldsymbol{a}}_{1}$ bond, the $b$ bond corresponds
to the ${\boldsymbol{a}}_{2}$ bond, and the $c$ bond corresponds
to the ${\boldsymbol{a}}_{3}$ bond. The couplings in the above equation
are related to the ones in the main context as follows,
\begin{eqnarray}
J & = & \frac{4}{3}J_{\pm}-\frac{2}{3}J_{\pm\pm}+\frac{1}{3}J_{zz},\\
K & = & 2J_{\pm\pm},\\
\Gamma & = & -\frac{2}{3}J_{\pm}-\frac{2}{3}J_{\pm\pm}+\frac{1}{3}J_{zz}. 
\end{eqnarray}

\section{Definition of effective spin-1/2 operators for the non-Kramers doublet}
\label{appendix2}

To give a simple illustration about the property of the non-Kramers doublet, we discuss the case for the spin-1 local moment ${\boldsymbol J}$. Assuming a single-ion anisotropy $-D_z (J^z)^2$ (with $D_z >0$), the three spin states are splitted into two lower doublets $J^z=\pm 1$ and an upper singlet $J^z=0$. The energy levels are shown in Fig.~\ref{sfig1}.
The $J^z = \pm 1$ states can be thought as a non-Kramers doublet. We thus define the effective spin-1/2 operators ${\boldsymbol S}$ from the physical spin operators ${\boldsymbol J}$, such that
\begin{eqnarray}
    S^z &=& \frac{1}{2} P J^z P , \\
    S^\pm &=& \frac{1}{2} P  \left(J^\pm\right)^2 P ,
\end{eqnarray}
where ${S^{\pm} = S^x \pm i S^y}$, ${J^{\pm} = J^x \pm i J^y}$, 
and $P$ is the projection operator onto the ${J^z=\pm 1}$ subspace. 
The effective spin operators defined above satisfy the canonical 
commutation relation,
\begin{eqnarray}
    [S^a, S^b] = i \epsilon_{abc} S^c.
\end{eqnarray}

We further identify $S^z$ as the dipolar moment and $S^{x,y}$ as the quadrupolar moment by examining their transformation under time reversal operation,
\begin{eqnarray}
    \Theta^{-1} S^z \Theta &=& -S^z,\\
    \Theta^{-1} S^x \Theta &=& S^x,\\
    \Theta^{-1} S^y \Theta &=& S^y.
\end{eqnarray}

\section{Spin-wave dispersion}
\label{appendix3}

In this Appendix, we plot the dispersion of all spin-wave branches. 
Compared with Fig.~\ref{fig6}, we see in Fig.~\ref{sfig2} 
that only some of the bands are visible in neutron scattering experiments.

\bibliography{refs}

\begin{thebibliography}{51}%
\makeatletter
\providecommand \@ifxundefined [1]{%
 \@ifx{#1\undefined}
}%
\providecommand \@ifnum [1]{%
 \ifnum #1\expandafter \@firstoftwo
 \else \expandafter \@secondoftwo
 \fi
}%
\providecommand \@ifx [1]{%
 \ifx #1\expandafter \@firstoftwo
 \else \expandafter \@secondoftwo
 \fi
}%
\providecommand \natexlab [1]{#1}%
\providecommand \enquote  [1]{``#1''}%
\providecommand \bibnamefont  [1]{#1}%
\providecommand \bibfnamefont [1]{#1}%
\providecommand \citenamefont [1]{#1}%
\providecommand \href@noop [0]{\@secondoftwo}%
\providecommand \href [0]{\begingroup \@sanitize@url \@href}%
\providecommand \@href[1]{\@@startlink{#1}\@@href}%
\providecommand \@@href[1]{\endgroup#1\@@endlink}%
\providecommand \@sanitize@url [0]{\catcode `\\12\catcode `\$12\catcode
  `\&12\catcode `\#12\catcode `\^12\catcode `\_12\catcode `\%12\relax}%
\providecommand \@@startlink[1]{}%
\providecommand \@@endlink[0]{}%
\providecommand \url  [0]{\begingroup\@sanitize@url \@url }%
\providecommand \@url [1]{\endgroup\@href {#1}{\urlprefix }}%
\providecommand \urlprefix  [0]{URL }%
\providecommand \Eprint [0]{\href }%
\providecommand \doibase [0]{http://dx.doi.org/}%
\providecommand \selectlanguage [0]{\@gobble}%
\providecommand \bibinfo  [0]{\@secondoftwo}%
\providecommand \bibfield  [0]{\@secondoftwo}%
\providecommand \translation [1]{[#1]}%
\providecommand \BibitemOpen [0]{}%
\providecommand \bibitemStop [0]{}%
\providecommand \bibitemNoStop [0]{.\EOS\space}%
\providecommand \EOS [0]{\spacefactor3000\relax}%
\providecommand \BibitemShut  [1]{\csname bibitem#1\endcsname}%
\let\auto@bib@innerbib\@empty
\bibitem [{\citenamefont {Witczak-Krempa}\ \emph {et~al.}(2014)\citenamefont
  {Witczak-Krempa}, \citenamefont {Chen}, \citenamefont {Kim},\ and\
  \citenamefont {Balents}}]{WCKB}%
  \BibitemOpen
  \bibfield  {author} {\bibinfo {author} {\bibfnamefont {William}\ \bibnamefont
  {Witczak-Krempa}}, \bibinfo {author} {\bibfnamefont {Gang}\ \bibnamefont
  {Chen}}, \bibinfo {author} {\bibfnamefont {Yong~Baek}\ \bibnamefont {Kim}}, \
  and\ \bibinfo {author} {\bibfnamefont {Leon}\ \bibnamefont {Balents}},\
  }\bibfield  {title} {\enquote {\bibinfo {title} {{Correlated Quantum
  Phenomena in the Strong Spin-Orbit Regime}},}\ }\href {\doibase
  10.1146/annurev-conmatphys-020911-125138} {\bibfield  {journal} {\bibinfo
  {journal} {Annual Review of Condensed Matter Physics}\ }\textbf {\bibinfo
  {volume} {5}},\ \bibinfo {pages} {57--82} (\bibinfo {year}
  {2014})}\BibitemShut {NoStop}%
\bibitem [{\citenamefont {Li}\ \emph {et~al.}(2015{\natexlab{a}})\citenamefont
  {Li}, \citenamefont {Liao}, \citenamefont {Zhang}, \citenamefont {Li},
  \citenamefont {Jin}, \citenamefont {Ling}, \citenamefont {Zhang},
  \citenamefont {Zou}, \citenamefont {Pi}, \citenamefont {Yang}, \citenamefont
  {Wang}, \citenamefont {Wu},\ and\ \citenamefont {Zhang}}]{srep}%
  \BibitemOpen
  \bibfield  {author} {\bibinfo {author} {\bibfnamefont {Yuesheng}\
  \bibnamefont {Li}}, \bibinfo {author} {\bibfnamefont {Haijun}\ \bibnamefont
  {Liao}}, \bibinfo {author} {\bibfnamefont {Zhen}\ \bibnamefont {Zhang}},
  \bibinfo {author} {\bibfnamefont {Shiyan}\ \bibnamefont {Li}}, \bibinfo
  {author} {\bibfnamefont {Feng}\ \bibnamefont {Jin}}, \bibinfo {author}
  {\bibfnamefont {Langsheng}\ \bibnamefont {Ling}}, \bibinfo {author}
  {\bibfnamefont {Lei}\ \bibnamefont {Zhang}}, \bibinfo {author} {\bibfnamefont
  {Youming}\ \bibnamefont {Zou}}, \bibinfo {author} {\bibfnamefont
  {Li}~\bibnamefont {Pi}}, \bibinfo {author} {\bibfnamefont {Zhaorong}\
  \bibnamefont {Yang}}, \bibinfo {author} {\bibfnamefont {Junfeng}\
  \bibnamefont {Wang}}, \bibinfo {author} {\bibfnamefont {Zhonghua}\
  \bibnamefont {Wu}}, \ and\ \bibinfo {author} {\bibfnamefont {Qingming}\
  \bibnamefont {Zhang}},\ }\bibfield  {title} {\enquote {\bibinfo {title}
  {{Gapless quantum spin liquid ground state in the two-dimensional spin-1/2
  triangular antiferromagnet YbMgGaO$_4$}},}\ }\href {\doibase
  10.1038/srep16419} {\bibfield  {journal} {\bibinfo  {journal} {Scientific
  Reports}\ }\textbf {\bibinfo {volume} {5}},\ \bibinfo {pages} {16419}
  (\bibinfo {year} {2015}{\natexlab{a}})}\BibitemShut {NoStop}%
\bibitem [{\citenamefont {Li}\ \emph {et~al.}(2015{\natexlab{b}})\citenamefont
  {Li}, \citenamefont {Chen}, \citenamefont {Tong}, \citenamefont {Pi},
  \citenamefont {Liu}, \citenamefont {Yang}, \citenamefont {Wang},\ and\
  \citenamefont {Zhang}}]{YueshengPRL}%
  \BibitemOpen
  \bibfield  {author} {\bibinfo {author} {\bibfnamefont {Yuesheng}\
  \bibnamefont {Li}}, \bibinfo {author} {\bibfnamefont {Gang}\ \bibnamefont
  {Chen}}, \bibinfo {author} {\bibfnamefont {Wei}\ \bibnamefont {Tong}},
  \bibinfo {author} {\bibfnamefont {Li}~\bibnamefont {Pi}}, \bibinfo {author}
  {\bibfnamefont {Juanjuan}\ \bibnamefont {Liu}}, \bibinfo {author}
  {\bibfnamefont {Zhaorong}\ \bibnamefont {Yang}}, \bibinfo {author}
  {\bibfnamefont {Xiaoqun}\ \bibnamefont {Wang}}, \ and\ \bibinfo {author}
  {\bibfnamefont {Qingming}\ \bibnamefont {Zhang}},\ }\bibfield  {title}
  {\enquote {\bibinfo {title} {{Rare-Earth Triangular Lattice Spin Liquid: A
  Single-Crystal Study of ${\mathrm{YbMgGaO}}_{4}$}},}\ }\href {\doibase
  10.1103/PhysRevLett.115.167203} {\bibfield  {journal} {\bibinfo  {journal}
  {Phys. Rev. Lett.}\ }\textbf {\bibinfo {volume} {115}},\ \bibinfo {pages}
  {167203} (\bibinfo {year} {2015}{\natexlab{b}})}\BibitemShut {NoStop}%
\bibitem [{\citenamefont {Li}\ \emph {et~al.}(2016{\natexlab{a}})\citenamefont
  {Li}, \citenamefont {Wang},\ and\ \citenamefont {Chen}}]{Yaodong2016PRB}%
  \BibitemOpen
  \bibfield  {author} {\bibinfo {author} {\bibfnamefont {Yao-Dong}\
  \bibnamefont {Li}}, \bibinfo {author} {\bibfnamefont {Xiaoqun}\ \bibnamefont
  {Wang}}, \ and\ \bibinfo {author} {\bibfnamefont {Gang}\ \bibnamefont
  {Chen}},\ }\bibfield  {title} {\enquote {\bibinfo {title} {Anisotropic spin
  model of strong spin-orbit-coupled triangular antiferromagnets},}\ }\href
  {\doibase 10.1103/PhysRevB.94.035107} {\bibfield  {journal} {\bibinfo
  {journal} {Phys. Rev. B}\ }\textbf {\bibinfo {volume} {94}},\ \bibinfo
  {pages} {035107} (\bibinfo {year} {2016}{\natexlab{a}})}\BibitemShut
  {NoStop}%
\bibitem [{\citenamefont {Shen}\ \emph {et~al.}(2016)\citenamefont {Shen},
  \citenamefont {Li}, \citenamefont {Wo}, \citenamefont {Li}, \citenamefont
  {Shen}, \citenamefont {Pan}, \citenamefont {Wang}, \citenamefont {Walker},
  \citenamefont {Steffens}, \citenamefont {Boehm}, \citenamefont {Hao},
  \citenamefont {Quintero-Castro}, \citenamefont {Harriger}, \citenamefont
  {Hao}, \citenamefont {Meng}, \citenamefont {Zhang}, \citenamefont {Chen},\
  and\ \citenamefont {Zhao}}]{YaoShenNature}%
  \BibitemOpen
  \bibfield  {author} {\bibinfo {author} {\bibfnamefont {Yao}\ \bibnamefont
  {Shen}}, \bibinfo {author} {\bibfnamefont {Yao-Dong}\ \bibnamefont {Li}},
  \bibinfo {author} {\bibfnamefont {Hongliang}\ \bibnamefont {Wo}}, \bibinfo
  {author} {\bibfnamefont {Yuesheng}\ \bibnamefont {Li}}, \bibinfo {author}
  {\bibfnamefont {Shoudong}\ \bibnamefont {Shen}}, \bibinfo {author}
  {\bibfnamefont {Bingying}\ \bibnamefont {Pan}}, \bibinfo {author}
  {\bibfnamefont {Qisi}\ \bibnamefont {Wang}}, \bibinfo {author} {\bibfnamefont
  {H.~C.}\ \bibnamefont {Walker}}, \bibinfo {author} {\bibfnamefont
  {P.}~\bibnamefont {Steffens}}, \bibinfo {author} {\bibfnamefont
  {M}~\bibnamefont {Boehm}}, \bibinfo {author} {\bibfnamefont {Yiqing}\
  \bibnamefont {Hao}}, \bibinfo {author} {\bibfnamefont {D.~L.}\ \bibnamefont
  {Quintero-Castro}}, \bibinfo {author} {\bibfnamefont {L.~W.}\ \bibnamefont
  {Harriger}}, \bibinfo {author} {\bibfnamefont {Lijie}\ \bibnamefont {Hao}},
  \bibinfo {author} {\bibfnamefont {Siqin}\ \bibnamefont {Meng}}, \bibinfo
  {author} {\bibfnamefont {Qingming}\ \bibnamefont {Zhang}}, \bibinfo {author}
  {\bibfnamefont {Gang}\ \bibnamefont {Chen}}, \ and\ \bibinfo {author}
  {\bibfnamefont {Jun}\ \bibnamefont {Zhao}},\ }\bibfield  {title} {\enquote
  {\bibinfo {title} {{Spinon Fermi surface in a triangular lattice quantum spin
  liquid {YbMgGaO$_4$}}},}\ }\href {\doibase 10.1038/nature20614} {\bibfield
  {journal} {\bibinfo  {journal} {Nature}\ }\textbf {\bibinfo {volume} {540}},\
  \bibinfo {pages} {559--562} (\bibinfo {year} {2016})}\BibitemShut {NoStop}%
\bibitem [{\citenamefont {Paddison}\ \emph {et~al.}(2017)\citenamefont
  {Paddison}, \citenamefont {Dun}, \citenamefont {Ehlers}, \citenamefont {Liu},
  \citenamefont {Stone}, \citenamefont {Zhou},\ and\ \citenamefont
  {Mourigal}}]{Martin2016}%
  \BibitemOpen
  \bibfield  {author} {\bibinfo {author} {\bibfnamefont {Joseph A.~M.}\
  \bibnamefont {Paddison}}, \bibinfo {author} {\bibfnamefont {Zhiling}\
  \bibnamefont {Dun}}, \bibinfo {author} {\bibfnamefont {Georg}\ \bibnamefont
  {Ehlers}}, \bibinfo {author} {\bibfnamefont {Yaohua}\ \bibnamefont {Liu}},
  \bibinfo {author} {\bibfnamefont {Matthew~B.}\ \bibnamefont {Stone}},
  \bibinfo {author} {\bibfnamefont {Haidong}\ \bibnamefont {Zhou}}, \ and\
  \bibinfo {author} {\bibfnamefont {Martin}\ \bibnamefont {Mourigal}},\
  }\bibfield  {title} {\enquote {\bibinfo {title} {Continuous excitations of
  the triangular-lattice quantum spin liquid {YbMgGaO$_4$}},}\ }\href {\doibase
  10.1038/NPHYS3971} {\bibfield  {journal} {\bibinfo  {journal} {Nature
  Physics}\ }\textbf {\bibinfo {volume} {13}},\ \bibinfo {pages} {117--122}
  (\bibinfo {year} {2017})}\BibitemShut {NoStop}%
\bibitem [{\citenamefont {Li}\ \emph {et~al.}(2018{\natexlab{a}})\citenamefont
  {Li}, \citenamefont {Shen}, \citenamefont {Li}, \citenamefont {Zhao},\ and\
  \citenamefont {Chen}}]{PhysRevB.97.125105}%
  \BibitemOpen
  \bibfield  {author} {\bibinfo {author} {\bibfnamefont {Yao-Dong}\
  \bibnamefont {Li}}, \bibinfo {author} {\bibfnamefont {Yao}\ \bibnamefont
  {Shen}}, \bibinfo {author} {\bibfnamefont {Yuesheng}\ \bibnamefont {Li}},
  \bibinfo {author} {\bibfnamefont {Jun}\ \bibnamefont {Zhao}}, \ and\ \bibinfo
  {author} {\bibfnamefont {Gang}\ \bibnamefont {Chen}},\ }\bibfield  {title}
  {\enquote {\bibinfo {title} {{Effect of spin-orbit coupling on the
  effective-spin correlation in ${\mathrm{YbMgGaO}}_{4}$}},}\ }\href {\doibase
  10.1103/PhysRevB.97.125105} {\bibfield  {journal} {\bibinfo  {journal} {Phys.
  Rev. B}\ }\textbf {\bibinfo {volume} {97}},\ \bibinfo {pages} {125105}
  (\bibinfo {year} {2018}{\natexlab{a}})}\BibitemShut {NoStop}%
\bibitem [{\citenamefont {Li}\ \emph {et~al.}(2017{\natexlab{a}})\citenamefont
  {Li}, \citenamefont {Lu},\ and\ \citenamefont {Chen}}]{PhysRevB.96.054445}%
  \BibitemOpen
  \bibfield  {author} {\bibinfo {author} {\bibfnamefont {Yao-Dong}\
  \bibnamefont {Li}}, \bibinfo {author} {\bibfnamefont {Yuan-Ming}\
  \bibnamefont {Lu}}, \ and\ \bibinfo {author} {\bibfnamefont {Gang}\
  \bibnamefont {Chen}},\ }\bibfield  {title} {\enquote {\bibinfo {title}
  {{Spinon Fermi surface $U(1)$ spin liquid in the spin-orbit-coupled
  triangular-lattice Mott insulator ${\mathrm{YbMgGaO}}_{4}$}},}\ }\href
  {\doibase 10.1103/PhysRevB.96.054445} {\bibfield  {journal} {\bibinfo
  {journal} {Phys. Rev. B}\ }\textbf {\bibinfo {volume} {96}},\ \bibinfo
  {pages} {054445} (\bibinfo {year} {2017}{\natexlab{a}})}\BibitemShut
  {NoStop}%
\bibitem [{\citenamefont {Li}\ \emph {et~al.}(2016{\natexlab{b}})\citenamefont
  {Li}, \citenamefont {Adroja}, \citenamefont {Biswas}, \citenamefont {Baker},
  \citenamefont {Zhang}, \citenamefont {Liu}, \citenamefont {Tsirlin},
  \citenamefont {Gegenwart},\ and\ \citenamefont {Zhang}}]{YueshengmuSR}%
  \BibitemOpen
  \bibfield  {author} {\bibinfo {author} {\bibfnamefont {Yuesheng}\
  \bibnamefont {Li}}, \bibinfo {author} {\bibfnamefont {Devashibhai}\
  \bibnamefont {Adroja}}, \bibinfo {author} {\bibfnamefont {Pabitra~K.}\
  \bibnamefont {Biswas}}, \bibinfo {author} {\bibfnamefont {Peter~J.}\
  \bibnamefont {Baker}}, \bibinfo {author} {\bibfnamefont {Qian}\ \bibnamefont
  {Zhang}}, \bibinfo {author} {\bibfnamefont {Juanjuan}\ \bibnamefont {Liu}},
  \bibinfo {author} {\bibfnamefont {Alexander~A.}\ \bibnamefont {Tsirlin}},
  \bibinfo {author} {\bibfnamefont {Philipp}\ \bibnamefont {Gegenwart}}, \ and\
  \bibinfo {author} {\bibfnamefont {Qingming}\ \bibnamefont {Zhang}},\
  }\bibfield  {title} {\enquote {\bibinfo {title} {{Muon Spin Relaxation
  Evidence for the U(1) Quantum Spin-Liquid Ground State in the Triangular
  Antiferromagnet ${\mathrm{YbMgGaO}}_{4}$}},}\ }\href {\doibase
  10.1103/PhysRevLett.117.097201} {\bibfield  {journal} {\bibinfo  {journal}
  {Phys. Rev. Lett.}\ }\textbf {\bibinfo {volume} {117}},\ \bibinfo {pages}
  {097201} (\bibinfo {year} {2016}{\natexlab{b}})}\BibitemShut {NoStop}%
\bibitem [{\citenamefont {Li}\ \emph {et~al.}(2016{\natexlab{c}})\citenamefont
  {Li}, \citenamefont {Wang},\ and\ \citenamefont {Chen}}]{Yaodong2016PRB2}%
  \BibitemOpen
  \bibfield  {author} {\bibinfo {author} {\bibfnamefont {Yao-Dong}\
  \bibnamefont {Li}}, \bibinfo {author} {\bibfnamefont {Xiaoqun}\ \bibnamefont
  {Wang}}, \ and\ \bibinfo {author} {\bibfnamefont {Gang}\ \bibnamefont
  {Chen}},\ }\bibfield  {title} {\enquote {\bibinfo {title} {Hidden multipolar
  orders of dipole-octupole doublets on a triangular lattice},}\ }\href
  {\doibase 10.1103/PhysRevB.94.201114} {\bibfield  {journal} {\bibinfo
  {journal} {Phys. Rev. B}\ }\textbf {\bibinfo {volume} {94}},\ \bibinfo
  {pages} {201114} (\bibinfo {year} {2016}{\natexlab{c}})}\BibitemShut
  {NoStop}%
\bibitem [{\citenamefont {Liu}\ \emph {et~al.}(2016)\citenamefont {Liu},
  \citenamefont {Yu},\ and\ \citenamefont {Wang}}]{ChangPRB}%
  \BibitemOpen
  \bibfield  {author} {\bibinfo {author} {\bibfnamefont {Changle}\ \bibnamefont
  {Liu}}, \bibinfo {author} {\bibfnamefont {Rong}\ \bibnamefont {Yu}}, \ and\
  \bibinfo {author} {\bibfnamefont {Xiaoqun}\ \bibnamefont {Wang}},\ }\bibfield
   {title} {\enquote {\bibinfo {title} {Semiclassical ground-state phase
  diagram and $\text{multi-}q$ phase of a spin-orbit-coupled model on
  triangular lattice},}\ }\href {\doibase 10.1103/PhysRevB.94.174424}
  {\bibfield  {journal} {\bibinfo  {journal} {Phys. Rev. B}\ }\textbf {\bibinfo
  {volume} {94}},\ \bibinfo {pages} {174424} (\bibinfo {year}
  {2016})}\BibitemShut {NoStop}%
\bibitem [{\citenamefont {Luo}\ \emph {et~al.}(2017)\citenamefont {Luo},
  \citenamefont {Hu}, \citenamefont {Xi}, \citenamefont {Zhao},\ and\
  \citenamefont {Wang}}]{PhysRevB.95.165110}%
  \BibitemOpen
  \bibfield  {author} {\bibinfo {author} {\bibfnamefont {Qiang}\ \bibnamefont
  {Luo}}, \bibinfo {author} {\bibfnamefont {Shijie}\ \bibnamefont {Hu}},
  \bibinfo {author} {\bibfnamefont {Bin}\ \bibnamefont {Xi}}, \bibinfo {author}
  {\bibfnamefont {Jize}\ \bibnamefont {Zhao}}, \ and\ \bibinfo {author}
  {\bibfnamefont {Xiaoqun}\ \bibnamefont {Wang}},\ }\bibfield  {title}
  {\enquote {\bibinfo {title} {{Ground-state phase diagram of an anisotropic
  spin-$\frac{1}{2}$ model on the triangular lattice}},}\ }\href {\doibase
  10.1103/PhysRevB.95.165110} {\bibfield  {journal} {\bibinfo  {journal} {Phys.
  Rev. B}\ }\textbf {\bibinfo {volume} {95}},\ \bibinfo {pages} {165110}
  (\bibinfo {year} {2017})}\BibitemShut {NoStop}%
\bibitem [{\citenamefont {Ma}\ \emph {et~al.}(2018)\citenamefont {Ma},
  \citenamefont {Wang}, \citenamefont {Dong}, \citenamefont {Zhang},
  \citenamefont {Li}, \citenamefont {Zheng}, \citenamefont {Yu}, \citenamefont
  {Wang}, \citenamefont {Che}, \citenamefont {Ran}, \citenamefont {Bao},
  \citenamefont {Cai}, \citenamefont {\ifmmode~\check{C}\else
  \v{C}\fi{}erm\'ak}, \citenamefont {Schneidewind}, \citenamefont {Yano},
  \citenamefont {Gardner}, \citenamefont {Lu}, \citenamefont {Yu},
  \citenamefont {Liu}, \citenamefont {Li}, \citenamefont {Li},\ and\
  \citenamefont {Wen}}]{PhysRevLett.120.087201}%
  \BibitemOpen
  \bibfield  {author} {\bibinfo {author} {\bibfnamefont {Zhen}\ \bibnamefont
  {Ma}}, \bibinfo {author} {\bibfnamefont {Jinghui}\ \bibnamefont {Wang}},
  \bibinfo {author} {\bibfnamefont {Zhao-Yang}\ \bibnamefont {Dong}}, \bibinfo
  {author} {\bibfnamefont {Jun}\ \bibnamefont {Zhang}}, \bibinfo {author}
  {\bibfnamefont {Shichao}\ \bibnamefont {Li}}, \bibinfo {author}
  {\bibfnamefont {Shu-Han}\ \bibnamefont {Zheng}}, \bibinfo {author}
  {\bibfnamefont {Yunjie}\ \bibnamefont {Yu}}, \bibinfo {author} {\bibfnamefont
  {Wei}\ \bibnamefont {Wang}}, \bibinfo {author} {\bibfnamefont {Liqiang}\
  \bibnamefont {Che}}, \bibinfo {author} {\bibfnamefont {Kejing}\ \bibnamefont
  {Ran}}, \bibinfo {author} {\bibfnamefont {Song}\ \bibnamefont {Bao}},
  \bibinfo {author} {\bibfnamefont {Zhengwei}\ \bibnamefont {Cai}}, \bibinfo
  {author} {\bibfnamefont {P.}~\bibnamefont {\ifmmode~\check{C}\else
  \v{C}\fi{}erm\'ak}}, \bibinfo {author} {\bibfnamefont {A.}~\bibnamefont
  {Schneidewind}}, \bibinfo {author} {\bibfnamefont {S.}~\bibnamefont {Yano}},
  \bibinfo {author} {\bibfnamefont {J.~S.}\ \bibnamefont {Gardner}}, \bibinfo
  {author} {\bibfnamefont {Xin}\ \bibnamefont {Lu}}, \bibinfo {author}
  {\bibfnamefont {Shun-Li}\ \bibnamefont {Yu}}, \bibinfo {author}
  {\bibfnamefont {Jun-Ming}\ \bibnamefont {Liu}}, \bibinfo {author}
  {\bibfnamefont {Shiyan}\ \bibnamefont {Li}}, \bibinfo {author} {\bibfnamefont
  {Jian-Xin}\ \bibnamefont {Li}}, \ and\ \bibinfo {author} {\bibfnamefont
  {Jinsheng}\ \bibnamefont {Wen}},\ }\bibfield  {title} {\enquote {\bibinfo
  {title} {{Spin-Glass Ground State in a Triangular-Lattice Compound
  ${\mathrm{YbZnGaO}}_{4}$}},}\ }\href {\doibase
  10.1103/PhysRevLett.120.087201} {\bibfield  {journal} {\bibinfo  {journal}
  {Phys. Rev. Lett.}\ }\textbf {\bibinfo {volume} {120}},\ \bibinfo {pages}
  {087201} (\bibinfo {year} {2018})}\BibitemShut {NoStop}%
\bibitem [{\citenamefont {Parker}\ and\ \citenamefont
  {Balents}()}]{Parker2018}%
  \BibitemOpen
  \bibfield  {author} {\bibinfo {author} {\bibfnamefont {Edward}\ \bibnamefont
  {Parker}}\ and\ \bibinfo {author} {\bibfnamefont {Leon}\ \bibnamefont
  {Balents}},\ }\bibfield  {title} {\enquote {\bibinfo {title}
  {{Finite-temperature behavior of a classical spin-orbit-coupled model for
  $\textrm{YbMgGaO}_4$ with and without bond disorder}},}\ }\href@noop {}
  {\bibfield  {journal} {\bibinfo  {journal} {arXiv}\ }\textbf {\bibinfo
  {volume} {1801.06941}}}\BibitemShut {NoStop}%
\bibitem [{\citenamefont {Xu}\ \emph {et~al.}(2016)\citenamefont {Xu},
  \citenamefont {Zhang}, \citenamefont {Li}, \citenamefont {Yu}, \citenamefont
  {Hong}, \citenamefont {Zhang},\ and\ \citenamefont {Li}}]{Shiyan2016}%
  \BibitemOpen
  \bibfield  {author} {\bibinfo {author} {\bibfnamefont {Y.}~\bibnamefont
  {Xu}}, \bibinfo {author} {\bibfnamefont {J.}~\bibnamefont {Zhang}}, \bibinfo
  {author} {\bibfnamefont {Y.~S.}\ \bibnamefont {Li}}, \bibinfo {author}
  {\bibfnamefont {Y.~J.}\ \bibnamefont {Yu}}, \bibinfo {author} {\bibfnamefont
  {X.~C.}\ \bibnamefont {Hong}}, \bibinfo {author} {\bibfnamefont {Q.~M.}\
  \bibnamefont {Zhang}}, \ and\ \bibinfo {author} {\bibfnamefont {S.~Y.}\
  \bibnamefont {Li}},\ }\bibfield  {title} {\enquote {\bibinfo {title}
  {{Absence of Magnetic Thermal Conductivity in the Quantum Spin-Liquid
  Candidate ${\mathrm{YbMgGaO}}_{4}$}},}\ }\href {\doibase
  10.1103/PhysRevLett.117.267202} {\bibfield  {journal} {\bibinfo  {journal}
  {Phys. Rev. Lett.}\ }\textbf {\bibinfo {volume} {117}},\ \bibinfo {pages}
  {267202} (\bibinfo {year} {2016})}\BibitemShut {NoStop}%
\bibitem [{\citenamefont {Toth}\ \emph {et~al.}(2017)\citenamefont {Toth},
  \citenamefont {Rolfs}, \citenamefont {Wildes},\ and\ \citenamefont
  {Ruegg}}]{Toth1705}%
  \BibitemOpen
  \bibfield  {author} {\bibinfo {author} {\bibfnamefont {Sandor}\ \bibnamefont
  {Toth}}, \bibinfo {author} {\bibfnamefont {Katharina}\ \bibnamefont {Rolfs}},
  \bibinfo {author} {\bibfnamefont {Andrew~R.}\ \bibnamefont {Wildes}}, \ and\
  \bibinfo {author} {\bibfnamefont {Christian}\ \bibnamefont {Ruegg}},\
  }\bibfield  {title} {\enquote {\bibinfo {title} {{Strong exchange anisotropy
  in YbMgGaO$_4$ from polarized neutron diffraction}},}\ }\href@noop {}
  {\bibfield  {journal} {\bibinfo  {journal} {arXiv preprint 1705.05699}\ }
  (\bibinfo {year} {2017})}\BibitemShut {NoStop}%
\bibitem [{\citenamefont {Yuesheng}\ \emph {et~al.}(2017)\citenamefont
  {Yuesheng}, \citenamefont {Adroja}, \citenamefont {Voneshen}, \citenamefont
  {Bewley}, \citenamefont {Zhang}, \citenamefont {Tsirlin},\ and\ \citenamefont
  {Gegenwart}}]{Yuesheng1704}%
  \BibitemOpen
  \bibfield  {author} {\bibinfo {author} {\bibfnamefont {Li}~\bibnamefont
  {Yuesheng}}, \bibinfo {author} {\bibfnamefont {Devashibhai}\ \bibnamefont
  {Adroja}}, \bibinfo {author} {\bibfnamefont {David}\ \bibnamefont
  {Voneshen}}, \bibinfo {author} {\bibfnamefont {Robert~I.}\ \bibnamefont
  {Bewley}}, \bibinfo {author} {\bibfnamefont {Qingming}\ \bibnamefont
  {Zhang}}, \bibinfo {author} {\bibfnamefont {Alexander~A.}\ \bibnamefont
  {Tsirlin}}, \ and\ \bibinfo {author} {\bibfnamefont {Philipp}\ \bibnamefont
  {Gegenwart}},\ }\bibfield  {title} {\enquote {\bibinfo {title}
  {{Nearest-neighbor resonating valence bonds in YbMgGaO$_4$}},}\ }\href
  {\doibase 10.1038/ncomms15814} {\bibfield  {journal} {\bibinfo  {journal}
  {Nature Communications, arXiv:1704.06468}\ }\textbf {\bibinfo {volume} {8}},\
  \bibinfo {pages} {15814} (\bibinfo {year} {2017})}\BibitemShut {NoStop}%
\bibitem [{\citenamefont {Li}\ \emph {et~al.}(2017{\natexlab{b}})\citenamefont
  {Li}, \citenamefont {Adroja}, \citenamefont {Bewley}, \citenamefont
  {Voneshen}, \citenamefont {Tsirlin}, \citenamefont {Gegenwart},\ and\
  \citenamefont {Zhang}}]{yuesheng1702}%
  \BibitemOpen
  \bibfield  {author} {\bibinfo {author} {\bibfnamefont {Yuesheng}\
  \bibnamefont {Li}}, \bibinfo {author} {\bibfnamefont {Devashibhai}\
  \bibnamefont {Adroja}}, \bibinfo {author} {\bibfnamefont {Robert~I.}\
  \bibnamefont {Bewley}}, \bibinfo {author} {\bibfnamefont {David}\
  \bibnamefont {Voneshen}}, \bibinfo {author} {\bibfnamefont {Alexander~A.}\
  \bibnamefont {Tsirlin}}, \bibinfo {author} {\bibfnamefont {Philipp}\
  \bibnamefont {Gegenwart}}, \ and\ \bibinfo {author} {\bibfnamefont
  {Qingming}\ \bibnamefont {Zhang}},\ }\bibfield  {title} {\enquote {\bibinfo
  {title} {{Crystalline Electric-Field Randomness in the Triangular Lattice
  Spin-Liquid ${\mathrm{YbMgGaO}}_{4}$}},}\ }\href {\doibase
  10.1103/PhysRevLett.118.107202} {\bibfield  {journal} {\bibinfo  {journal}
  {Phys. Rev. Lett.}\ }\textbf {\bibinfo {volume} {118}},\ \bibinfo {pages}
  {107202} (\bibinfo {year} {2017}{\natexlab{b}})}\BibitemShut {NoStop}%
\bibitem [{\citenamefont {Shen}\ \emph {et~al.}(2017)\citenamefont {Shen},
  \citenamefont {Li}, \citenamefont {Walker}, \citenamefont {Steffens},
  \citenamefont {Boehm}, \citenamefont {Zhang}, \citenamefont {Shen},
  \citenamefont {Wo}, \citenamefont {Chen},\ and\ \citenamefont
  {Zhao}}]{JunZhaounpub}%
  \BibitemOpen
  \bibfield  {author} {\bibinfo {author} {\bibfnamefont {Yao}\ \bibnamefont
  {Shen}}, \bibinfo {author} {\bibfnamefont {Yao-Dong}\ \bibnamefont {Li}},
  \bibinfo {author} {\bibfnamefont {H.~C.}\ \bibnamefont {Walker}}, \bibinfo
  {author} {\bibfnamefont {P.}~\bibnamefont {Steffens}}, \bibinfo {author}
  {\bibfnamefont {M.}~\bibnamefont {Boehm}}, \bibinfo {author} {\bibfnamefont
  {Xiaowen}\ \bibnamefont {Zhang}}, \bibinfo {author} {\bibfnamefont
  {Shoudong}\ \bibnamefont {Shen}}, \bibinfo {author} {\bibfnamefont
  {Hongliang}\ \bibnamefont {Wo}}, \bibinfo {author} {\bibfnamefont {Gang}\
  \bibnamefont {Chen}}, \ and\ \bibinfo {author} {\bibfnamefont {Jun}\
  \bibnamefont {Zhao}},\ }\bibfield  {title} {\enquote {\bibinfo {title}
  {{Fractionalized excitations in the partially magnetized spin liquid
  candidate YbMgGaO$_4$}},}\ }\href@noop {} {\bibfield  {journal} {\bibinfo
  {journal} {arXiv preprint arXiv:1708.06655}\ } (\bibinfo {year}
  {2017})}\BibitemShut {NoStop}%
\bibitem [{\citenamefont {Iaconis}\ \emph {et~al.}(2018)\citenamefont
  {Iaconis}, \citenamefont {Liu}, \citenamefont {Halász},\ and\ \citenamefont
  {Balents}}]{SciPostPhys.4.1.003}%
  \BibitemOpen
  \bibfield  {author} {\bibinfo {author} {\bibfnamefont {Jason}\ \bibnamefont
  {Iaconis}}, \bibinfo {author} {\bibfnamefont {Chunxiao}\ \bibnamefont {Liu}},
  \bibinfo {author} {\bibfnamefont {Gábor~B.}\ \bibnamefont {Halász}}, \ and\
  \bibinfo {author} {\bibfnamefont {Leon}\ \bibnamefont {Balents}},\ }\bibfield
   {title} {\enquote {\bibinfo {title} {{Spin Liquid versus Spin Orbit Coupling
  on the Triangular Lattice}},}\ }\href {\doibase 10.21468/SciPostPhys.4.1.003}
  {\bibfield  {journal} {\bibinfo  {journal} {SciPost Phys.}\ }\textbf
  {\bibinfo {volume} {4}},\ \bibinfo {pages} {003} (\bibinfo {year}
  {2018})}\BibitemShut {NoStop}%
\bibitem [{\citenamefont {Zhu}\ \emph {et~al.}(2017)\citenamefont {Zhu},
  \citenamefont {Maksimov}, \citenamefont {White},\ and\ \citenamefont
  {Chernyshev}}]{PhysRevLett.119.157201}%
  \BibitemOpen
  \bibfield  {author} {\bibinfo {author} {\bibfnamefont {Zhenyue}\ \bibnamefont
  {Zhu}}, \bibinfo {author} {\bibfnamefont {P.~A.}\ \bibnamefont {Maksimov}},
  \bibinfo {author} {\bibfnamefont {Steven~R.}\ \bibnamefont {White}}, \ and\
  \bibinfo {author} {\bibfnamefont {A.~L.}\ \bibnamefont {Chernyshev}},\
  }\bibfield  {title} {\enquote {\bibinfo {title} {{Disorder-Induced Mimicry of
  a Spin Liquid in ${\mathrm{YbMgGaO}}_{4}$}},}\ }\href {\doibase
  10.1103/PhysRevLett.119.157201} {\bibfield  {journal} {\bibinfo  {journal}
  {Phys. Rev. Lett.}\ }\textbf {\bibinfo {volume} {119}},\ \bibinfo {pages}
  {157201} (\bibinfo {year} {2017})}\BibitemShut {NoStop}%
\bibitem [{\citenamefont {Luo}\ \emph {et~al.}(2018)\citenamefont {Luo},
  \citenamefont {Lake}, \citenamefont {Mei},\ and\ \citenamefont
  {Starykh}}]{PhysRevLett.120.037204}%
  \BibitemOpen
  \bibfield  {author} {\bibinfo {author} {\bibfnamefont {Zhu-Xi}\ \bibnamefont
  {Luo}}, \bibinfo {author} {\bibfnamefont {Ethan}\ \bibnamefont {Lake}},
  \bibinfo {author} {\bibfnamefont {Jia-Wei}\ \bibnamefont {Mei}}, \ and\
  \bibinfo {author} {\bibfnamefont {Oleg~A.}\ \bibnamefont {Starykh}},\
  }\bibfield  {title} {\enquote {\bibinfo {title} {Spinon magnetic resonance of
  quantum spin liquids},}\ }\href {\doibase 10.1103/PhysRevLett.120.037204}
  {\bibfield  {journal} {\bibinfo  {journal} {Phys. Rev. Lett.}\ }\textbf
  {\bibinfo {volume} {120}},\ \bibinfo {pages} {037204} (\bibinfo {year}
  {2018})}\BibitemShut {NoStop}%
\bibitem [{\citenamefont {Kimchi}\ \emph {et~al.}(2017)\citenamefont {Kimchi},
  \citenamefont {Nahum},\ and\ \citenamefont {Senthil}}]{1710.06860}%
  \BibitemOpen
  \bibfield  {author} {\bibinfo {author} {\bibfnamefont {I}~\bibnamefont
  {Kimchi}}, \bibinfo {author} {\bibfnamefont {A.}~\bibnamefont {Nahum}}, \
  and\ \bibinfo {author} {\bibfnamefont {T.}~\bibnamefont {Senthil}},\
  }\bibfield  {title} {\enquote {\bibinfo {title} {{Valence Bonds in Random
  Quantum Magnets: Theory and Application to YbMgGaO$_4$}},}\ }\href@noop {}
  {\bibfield  {journal} {\bibinfo  {journal} {arXiv preprint 1710.06860}\ }
  (\bibinfo {year} {2017})}\BibitemShut {NoStop}%
\bibitem [{\citenamefont {Zhu}\ \emph {et~al.}(2018)\citenamefont {Zhu},
  \citenamefont {Maksimov}, \citenamefont {White},\ and\ \citenamefont
  {Chernyshev}}]{1801.01130}%
  \BibitemOpen
  \bibfield  {author} {\bibinfo {author} {\bibfnamefont {Zhenyue}\ \bibnamefont
  {Zhu}}, \bibinfo {author} {\bibfnamefont {P.A.}\ \bibnamefont {Maksimov}},
  \bibinfo {author} {\bibfnamefont {S.R.}\ \bibnamefont {White}}, \ and\
  \bibinfo {author} {\bibfnamefont {A.L.}\ \bibnamefont {Chernyshev}},\
  }\bibfield  {title} {\enquote {\bibinfo {title} {{Topography of Spin Liquids
  on a Triangular Lattice}},}\ }\href@noop {} {\bibfield  {journal} {\bibinfo
  {journal} {arXiv preprint 1801.01130}\ } (\bibinfo {year}
  {2018})}\BibitemShut {NoStop}%
\bibitem [{\citenamefont {{Stoyko}}\ and\ \citenamefont
  {{Mar}}(2011)}]{triangle1}%
  \BibitemOpen
  \bibfield  {author} {\bibinfo {author} {\bibfnamefont {S.~S.}\ \bibnamefont
  {{Stoyko}}}\ and\ \bibinfo {author} {\bibfnamefont {A.}~\bibnamefont
  {{Mar}}},\ }\bibfield  {title} {\enquote {\bibinfo {title} {{{Ternary
  rare-earth zinc arsenides REZn$_{ 1- x }$As$_{ 2 }$ ( RE =La-Nd, Sm)}}},}\
  }\href {\doibase 10.1016/j.jssc.2011.07.002} {\bibfield  {journal} {\bibinfo
  {journal} {Journal of Solid State Chemistry France}\ }\textbf {\bibinfo
  {volume} {184}},\ \bibinfo {pages} {2360--2367} (\bibinfo {year}
  {2011})}\BibitemShut {NoStop}%
\bibitem [{\citenamefont {{Nientiedt}}\ and\ \citenamefont
  {{Jeitschko}}(1999)}]{triangle2}%
  \BibitemOpen
  \bibfield  {author} {\bibinfo {author} {\bibfnamefont {A.~T.}\ \bibnamefont
  {{Nientiedt}}}\ and\ \bibinfo {author} {\bibfnamefont {W.}~\bibnamefont
  {{Jeitschko}}},\ }\bibfield  {title} {\enquote {\bibinfo {title} {{The Series
  of Rare Earth Zinc Phosphides RZn$_{3}$P$_{3}$ ( R=Y, La-Nd, Sm, Gd-Er) and
  the Corresponding Cadmium Compound PrCd$_{3}$P$_{3}$}},}\ }\href {\doibase
  10.1006/jssc.1999.8396} {\bibfield  {journal} {\bibinfo  {journal} {Journal
  of Solid State Chemistry France}\ }\textbf {\bibinfo {volume} {146}},\
  \bibinfo {pages} {478--483} (\bibinfo {year} {1999})}\BibitemShut {NoStop}%
\bibitem [{\citenamefont {{Yamada}}\ \emph {et~al.}(2010)\citenamefont
  {{Yamada}}, \citenamefont {{Hara}}, \citenamefont {{Matsubayashi}},
  \citenamefont {{Munakata}}, \citenamefont {{Ganguli}}, \citenamefont
  {{Ochiai}}, \citenamefont {{Matsumoto}},\ and\ \citenamefont
  {{Uwatoko}}}]{triangle3}%
  \BibitemOpen
  \bibfield  {author} {\bibinfo {author} {\bibfnamefont {A.}~\bibnamefont
  {{Yamada}}}, \bibinfo {author} {\bibfnamefont {N.}~\bibnamefont {{Hara}}},
  \bibinfo {author} {\bibfnamefont {K.}~\bibnamefont {{Matsubayashi}}},
  \bibinfo {author} {\bibfnamefont {K.}~\bibnamefont {{Munakata}}}, \bibinfo
  {author} {\bibfnamefont {C.}~\bibnamefont {{Ganguli}}}, \bibinfo {author}
  {\bibfnamefont {A.}~\bibnamefont {{Ochiai}}}, \bibinfo {author}
  {\bibfnamefont {T.}~\bibnamefont {{Matsumoto}}}, \ and\ \bibinfo {author}
  {\bibfnamefont {Y.}~\bibnamefont {{Uwatoko}}},\ }\bibfield  {title} {\enquote
  {\bibinfo {title} {{{Effect of pressure on the electrical resistivity of
  CeZn$_{3}$P$_{3}$}}},}\ }in\ \href {\doibase 10.1088/1742-6596/215/1/012031}
  {\emph {\bibinfo {booktitle} {Journal of Physics Conference Series}}},\
  \bibinfo {series} {Journal of Physics Conference Series}, Vol.\ \bibinfo
  {volume} {215}\ (\bibinfo {year} {2010})\ p.\ \bibinfo {pages}
  {012031}\BibitemShut {NoStop}%
\bibitem [{\citenamefont {Shevchenko}\ \emph {et~al.}(2017)\citenamefont
  {Shevchenko}, \citenamefont {Kononova}, \citenamefont {Kokh}, \citenamefont
  {Bolatov}, \citenamefont {Uralbekov},\ and\ \citenamefont
  {Burkitbaev}}]{triangle4}%
  \BibitemOpen
  \bibfield  {author} {\bibinfo {author} {\bibfnamefont {V.~S.}\ \bibnamefont
  {Shevchenko}}, \bibinfo {author} {\bibfnamefont {N.~G.}\ \bibnamefont
  {Kononova}}, \bibinfo {author} {\bibfnamefont {A.~E.}\ \bibnamefont {Kokh}},
  \bibinfo {author} {\bibfnamefont {A.~K.}\ \bibnamefont {Bolatov}}, \bibinfo
  {author} {\bibfnamefont {B.~M.}\ \bibnamefont {Uralbekov}}, \ and\ \bibinfo
  {author} {\bibfnamefont {M.~M.}\ \bibnamefont {Burkitbaev}},\ }\bibfield
  {title} {\enquote {\bibinfo {title} {{KBaR(BO$_3$)$_2$ orthoborates (R = RE):
  Synthesis and study}},}\ }\href {\doibase 10.1134/S0036023617090133}
  {\bibfield  {journal} {\bibinfo  {journal} {Russian Journal of Inorganic
  Chemistry}\ }\textbf {\bibinfo {volume} {62}},\ \bibinfo {pages} {1177--1181}
  (\bibinfo {year} {2017})}\BibitemShut {NoStop}%
\bibitem [{\citenamefont {Ohtani}\ \emph {et~al.}(1987)\citenamefont {Ohtani},
  \citenamefont {Honjo},\ and\ \citenamefont {Wada}}]{OHTANI}%
  \BibitemOpen
  \bibfield  {author} {\bibinfo {author} {\bibfnamefont {T.}~\bibnamefont
  {Ohtani}}, \bibinfo {author} {\bibfnamefont {H.}~\bibnamefont {Honjo}}, \
  and\ \bibinfo {author} {\bibfnamefont {H.}~\bibnamefont {Wada}},\ }\bibfield
  {title} {\enquote {\bibinfo {title} {{ Synthesis, Order-disorder transition
  and magnetic properties of LiLnS$_2$, LiLnSe$_2$, NaLnS$_2$ and NaLnSe$_2$
  (Lnflanthanides) }},}\ }\href@noop {} {\bibfield  {journal} {\bibinfo
  {journal} {Mat. Res. Bull.}\ }\textbf {\bibinfo {volume} {22}},\ \bibinfo
  {pages} {829--840} (\bibinfo {year} {1987})}\BibitemShut {NoStop}%
\bibitem [{\citenamefont {Sato}\ \emph {et~al.}(1984)\citenamefont {Sato},
  \citenamefont {Adachi},\ and\ \citenamefont {Shiokawa}}]{Sato}%
  \BibitemOpen
  \bibfield  {author} {\bibinfo {author} {\bibfnamefont {M.}~\bibnamefont
  {Sato}}, \bibinfo {author} {\bibfnamefont {G.}~\bibnamefont {Adachi}}, \ and\
  \bibinfo {author} {\bibfnamefont {J.}~\bibnamefont {Shiokawa}},\ }\bibfield
  {title} {\enquote {\bibinfo {title} {{Preparation and structure of sodium
  rare-earth sulfides, NaLnS$_2$ (Ln; Rare earth elements)}},}\ }\href@noop {}
  {\bibfield  {journal} {\bibinfo  {journal} {Mat. Res. Bull.}\ }\textbf
  {\bibinfo {volume} {19}},\ \bibinfo {pages} {1215--1220} (\bibinfo {year}
  {1984})}\BibitemShut {NoStop}%
\bibitem [{\citenamefont {Huang}\ \emph {et~al.}(2014)\citenamefont {Huang},
  \citenamefont {Chen},\ and\ \citenamefont
  {Hermele}}]{PhysRevLett.112.167203}%
  \BibitemOpen
  \bibfield  {author} {\bibinfo {author} {\bibfnamefont {Yi-Ping}\ \bibnamefont
  {Huang}}, \bibinfo {author} {\bibfnamefont {Gang}\ \bibnamefont {Chen}}, \
  and\ \bibinfo {author} {\bibfnamefont {Michael}\ \bibnamefont {Hermele}},\
  }\bibfield  {title} {\enquote {\bibinfo {title} {{Quantum Spin Ices and
  Topological Phases from Dipolar-Octupolar Doublets on the Pyrochlore
  Lattice}},}\ }\href {\doibase 10.1103/PhysRevLett.112.167203} {\bibfield
  {journal} {\bibinfo  {journal} {Phys. Rev. Lett.}\ }\textbf {\bibinfo
  {volume} {112}},\ \bibinfo {pages} {167203} (\bibinfo {year}
  {2014})}\BibitemShut {NoStop}%
\bibitem [{\citenamefont {Li}\ and\ \citenamefont
  {Chen}(2017)}]{PhysRevB.95.041106}%
  \BibitemOpen
  \bibfield  {author} {\bibinfo {author} {\bibfnamefont {Yao-Dong}\
  \bibnamefont {Li}}\ and\ \bibinfo {author} {\bibfnamefont {Gang}\
  \bibnamefont {Chen}},\ }\bibfield  {title} {\enquote {\bibinfo {title}
  {{Symmetry enriched U(1) topological orders for dipole-octupole doublets on a
  pyrochlore lattice}},}\ }\href {\doibase 10.1103/PhysRevB.95.041106}
  {\bibfield  {journal} {\bibinfo  {journal} {Phys. Rev. B}\ }\textbf {\bibinfo
  {volume} {95}},\ \bibinfo {pages} {041106} (\bibinfo {year}
  {2017})}\BibitemShut {NoStop}%
\bibitem [{\citenamefont {Lee}\ \emph {et~al.}(2012)\citenamefont {Lee},
  \citenamefont {Onoda},\ and\ \citenamefont {Balents}}]{PhysRevB.86.104412}%
  \BibitemOpen
  \bibfield  {author} {\bibinfo {author} {\bibfnamefont {SungBin}\ \bibnamefont
  {Lee}}, \bibinfo {author} {\bibfnamefont {Shigeki}\ \bibnamefont {Onoda}}, \
  and\ \bibinfo {author} {\bibfnamefont {Leon}\ \bibnamefont {Balents}},\
  }\bibfield  {title} {\enquote {\bibinfo {title} {Generic quantum spin ice},}\
  }\href {\doibase 10.1103/PhysRevB.86.104412} {\bibfield  {journal} {\bibinfo
  {journal} {Phys. Rev. B}\ }\textbf {\bibinfo {volume} {86}},\ \bibinfo
  {pages} {104412} (\bibinfo {year} {2012})}\BibitemShut {NoStop}%
\bibitem [{\citenamefont {Onoda}\ and\ \citenamefont
  {Tanaka}(2010)}]{PhysRevLett.105.047201}%
  \BibitemOpen
  \bibfield  {author} {\bibinfo {author} {\bibfnamefont {Shigeki}\ \bibnamefont
  {Onoda}}\ and\ \bibinfo {author} {\bibfnamefont {Yoichi}\ \bibnamefont
  {Tanaka}},\ }\bibfield  {title} {\enquote {\bibinfo {title} {{Quantum Melting
  of Spin Ice: Emergent Cooperative Quadrupole and Chirality}},}\ }\href
  {\doibase 10.1103/PhysRevLett.105.047201} {\bibfield  {journal} {\bibinfo
  {journal} {Phys. Rev. Lett.}\ }\textbf {\bibinfo {volume} {105}},\ \bibinfo
  {pages} {047201} (\bibinfo {year} {2010})}\BibitemShut {NoStop}%
\bibitem [{\citenamefont {Maharaj}\ \emph {et~al.}()\citenamefont {Maharaj},
  \citenamefont {Rosenberg}, \citenamefont {Hristov}, \citenamefont {Berg},
  \citenamefont {Fernandes}, \citenamefont {Fisher},\ and\ \citenamefont
  {Kivelson}}]{Kivelson}%
  \BibitemOpen
  \bibfield  {author} {\bibinfo {author} {\bibfnamefont {Akash~V.}\
  \bibnamefont {Maharaj}}, \bibinfo {author} {\bibfnamefont {Elliott~W.}\
  \bibnamefont {Rosenberg}}, \bibinfo {author} {\bibfnamefont {Alexander~T.}\
  \bibnamefont {Hristov}}, \bibinfo {author} {\bibfnamefont {Erez}\
  \bibnamefont {Berg}}, \bibinfo {author} {\bibfnamefont {Rafael~M.}\
  \bibnamefont {Fernandes}}, \bibinfo {author} {\bibfnamefont {Ian~R.}\
  \bibnamefont {Fisher}}, \ and\ \bibinfo {author} {\bibfnamefont {Steven~A.}\
  \bibnamefont {Kivelson}},\ }\bibfield  {title} {\enquote {\bibinfo {title}
  {{Transverse fields to tune an Ising-nematic quantum critical transition}},}\
  }\href@noop {} {\bibfield  {journal} {\bibinfo  {journal} {arXiv}\ }\textbf
  {\bibinfo {volume} {1705.01111}}}\BibitemShut {NoStop}%
\bibitem [{\citenamefont {Luttinger}\ and\ \citenamefont
  {Tisza}(1946)}]{LuttingerTisza}%
  \BibitemOpen
  \bibfield  {author} {\bibinfo {author} {\bibfnamefont {J.~M.}\ \bibnamefont
  {Luttinger}}\ and\ \bibinfo {author} {\bibfnamefont {L.}~\bibnamefont
  {Tisza}},\ }\bibfield  {title} {\enquote {\bibinfo {title} {Theory of dipole
  interaction in crystals},}\ }\href {\doibase 10.1103/PhysRev.70.954}
  {\bibfield  {journal} {\bibinfo  {journal} {Phys. Rev.}\ }\textbf {\bibinfo
  {volume} {70}},\ \bibinfo {pages} {954--964} (\bibinfo {year}
  {1946})}\BibitemShut {NoStop}%
\bibitem [{\citenamefont {Wessel}\ and\ \citenamefont
  {Troyer}(2005)}]{troyer2005supersolid}%
  \BibitemOpen
  \bibfield  {author} {\bibinfo {author} {\bibfnamefont {Stefan}\ \bibnamefont
  {Wessel}}\ and\ \bibinfo {author} {\bibfnamefont {Matthias}\ \bibnamefont
  {Troyer}},\ }\bibfield  {title} {\enquote {\bibinfo {title} {Supersolid
  hard-core bosons on the triangular lattice},}\ }\href {\doibase
  10.1103/PhysRevLett.95.127205} {\bibfield  {journal} {\bibinfo  {journal}
  {Phys. Rev. Lett.}\ }\textbf {\bibinfo {volume} {95}},\ \bibinfo {pages}
  {127205} (\bibinfo {year} {2005})}\BibitemShut {NoStop}%
\bibitem [{\citenamefont {Melko}\ \emph {et~al.}(2005)\citenamefont {Melko},
  \citenamefont {Paramekanti}, \citenamefont {Burkov}, \citenamefont
  {Vishwanath}, \citenamefont {Sheng},\ and\ \citenamefont
  {Balents}}]{balents2005supersolid}%
  \BibitemOpen
  \bibfield  {author} {\bibinfo {author} {\bibfnamefont {R.~G.}\ \bibnamefont
  {Melko}}, \bibinfo {author} {\bibfnamefont {A.}~\bibnamefont {Paramekanti}},
  \bibinfo {author} {\bibfnamefont {A.~A.}\ \bibnamefont {Burkov}}, \bibinfo
  {author} {\bibfnamefont {A.}~\bibnamefont {Vishwanath}}, \bibinfo {author}
  {\bibfnamefont {D.~N.}\ \bibnamefont {Sheng}}, \ and\ \bibinfo {author}
  {\bibfnamefont {L.}~\bibnamefont {Balents}},\ }\bibfield  {title} {\enquote
  {\bibinfo {title} {Supersolid order from disorder: Hard-core bosons on the
  triangular lattice},}\ }\href {\doibase 10.1103/PhysRevLett.95.127207}
  {\bibfield  {journal} {\bibinfo  {journal} {Phys. Rev. Lett.}\ }\textbf
  {\bibinfo {volume} {95}},\ \bibinfo {pages} {127207} (\bibinfo {year}
  {2005})}\BibitemShut {NoStop}%
\bibitem [{\citenamefont {Yamamoto}\ \emph {et~al.}(2014)\citenamefont
  {Yamamoto}, \citenamefont {Marmorini},\ and\ \citenamefont
  {Danshita}}]{danshita2014xxz}%
  \BibitemOpen
  \bibfield  {author} {\bibinfo {author} {\bibfnamefont {Daisuke}\ \bibnamefont
  {Yamamoto}}, \bibinfo {author} {\bibfnamefont {Giacomo}\ \bibnamefont
  {Marmorini}}, \ and\ \bibinfo {author} {\bibfnamefont {Ippei}\ \bibnamefont
  {Danshita}},\ }\bibfield  {title} {\enquote {\bibinfo {title} {{Quantum Phase
  Diagram of the Triangular-Lattice $XXZ$ Model in a Magnetic Field}},}\ }\href
  {\doibase 10.1103/PhysRevLett.112.127203} {\bibfield  {journal} {\bibinfo
  {journal} {Phys. Rev. Lett.}\ }\textbf {\bibinfo {volume} {112}},\ \bibinfo
  {pages} {127203} (\bibinfo {year} {2014})}\BibitemShut {NoStop}%
\bibitem [{\citenamefont {{Petit, S.}}(2011)}]{petit2011}%
  \BibitemOpen
  \bibfield  {author} {\bibinfo {author} {\bibnamefont {{Petit, S.}}},\
  }\bibfield  {title} {\enquote {\bibinfo {title} {Numerical simulations and
  magnetism},}\ }\href {\doibase 10.1051/sfn/201112006} {\bibfield  {journal}
  {\bibinfo  {journal} {JDN}\ }\textbf {\bibinfo {volume} {12}},\ \bibinfo
  {pages} {105--121} (\bibinfo {year} {2011})}\BibitemShut {NoStop}%
\bibitem [{\citenamefont {Wallace}(1962)}]{wallace1962}%
  \BibitemOpen
  \bibfield  {author} {\bibinfo {author} {\bibfnamefont {Duane~C.}\
  \bibnamefont {Wallace}},\ }\bibfield  {title} {\enquote {\bibinfo {title}
  {{Spin Waves in Complex Lattices}},}\ }\href {\doibase
  10.1103/PhysRev.128.1614} {\bibfield  {journal} {\bibinfo  {journal} {Phys.
  Rev.}\ }\textbf {\bibinfo {volume} {128}},\ \bibinfo {pages} {1614--1618}
  (\bibinfo {year} {1962})}\BibitemShut {NoStop}%
\bibitem [{\citenamefont {Toth}\ and\ \citenamefont {Lake}(2015)}]{toth2015}%
  \BibitemOpen
  \bibfield  {author} {\bibinfo {author} {\bibfnamefont {S}~\bibnamefont
  {Toth}}\ and\ \bibinfo {author} {\bibfnamefont {B}~\bibnamefont {Lake}},\
  }\bibfield  {title} {\enquote {\bibinfo {title} {{Linear spin wave theory for
  single-Q incommensurate magnetic structures}},}\ }\href
  {http://stacks.iop.org/0953-8984/27/i=16/a=166002} {\bibfield  {journal}
  {\bibinfo  {journal} {Journal of Physics: Condensed Matter}\ }\textbf
  {\bibinfo {volume} {27}},\ \bibinfo {pages} {166002} (\bibinfo {year}
  {2015})}\BibitemShut {NoStop}%
\bibitem [{\citenamefont {Goodman}\ and\ \citenamefont
  {Wallach}(2009)}]{Goodman2009}%
  \BibitemOpen
  \bibfield  {author} {\bibinfo {author} {\bibfnamefont {Roe}\ \bibnamefont
  {Goodman}}\ and\ \bibinfo {author} {\bibfnamefont {Nolan~R}\ \bibnamefont
  {Wallach}},\ }\href@noop {} {\emph {\bibinfo {title} {Symmetry,
  representations, and invariants}}},\ Vol.~\bibinfo {volume} {66}\ (\bibinfo
  {publisher} {Springer},\ \bibinfo {year} {2009})\BibitemShut {NoStop}%
\bibitem [{\citenamefont {{Cevallos}}\ \emph {et~al.}(2017)\citenamefont
  {{Cevallos}}, \citenamefont {{Stolze}}, \citenamefont {{Kong}},\ and\
  \citenamefont {{Cava}}}]{cavaTMGO}%
  \BibitemOpen
  \bibfield  {author} {\bibinfo {author} {\bibfnamefont {F.~A.}\ \bibnamefont
  {{Cevallos}}}, \bibinfo {author} {\bibfnamefont {K.}~\bibnamefont
  {{Stolze}}}, \bibinfo {author} {\bibfnamefont {T.}~\bibnamefont {{Kong}}}, \
  and\ \bibinfo {author} {\bibfnamefont {R.~J.}\ \bibnamefont {{Cava}}},\
  }\bibfield  {title} {\enquote {\bibinfo {title} {{{Anisotropic magnetic
  properties of the triangular plane lattice material TmMgGaO$_4$}}},}\
  }\href@noop {} {\bibfield  {journal} {\bibinfo  {journal} {ArXiv e-prints}\ }
  (\bibinfo {year} {2017})},\ \Eprint {http://arxiv.org/abs/1710.07707}
  {arXiv:1710.07707 [cond-mat.mtrl-sci]} \BibitemShut {NoStop}%
\bibitem [{\citenamefont {Li}\ \emph {et~al.}(2018{\natexlab{b}})\citenamefont
  {Li}, \citenamefont {Bachus}, \citenamefont {Tokiwa}, \citenamefont
  {Tsirlin},\ and\ \citenamefont {Gegenwart}}]{1804.00696}%
  \BibitemOpen
  \bibfield  {author} {\bibinfo {author} {\bibfnamefont {Yuesheng}\
  \bibnamefont {Li}}, \bibinfo {author} {\bibfnamefont {Sebastian}\
  \bibnamefont {Bachus}}, \bibinfo {author} {\bibfnamefont {Yoshifumi}\
  \bibnamefont {Tokiwa}}, \bibinfo {author} {\bibfnamefont {Alexander~A.}\
  \bibnamefont {Tsirlin}}, \ and\ \bibinfo {author} {\bibfnamefont {Philipp}\
  \bibnamefont {Gegenwart}},\ }\bibfield  {title} {\enquote {\bibinfo {title}
  {{Absence of zero-point entropy in a triangular Ising antiferromagnet}},}\
  }\href@noop {} {\bibfield  {journal} {\bibinfo  {journal} {arXiv preprint
  1804.00696}\ } (\bibinfo {year} {2018}{\natexlab{b}})}\BibitemShut {NoStop}%
\bibitem [{\citenamefont {Liu}\ and\ \citenamefont
  {Chen}(2018)}]{Changleunpub}%
  \BibitemOpen
  \bibfield  {author} {\bibinfo {author} {\bibfnamefont {Changle}\ \bibnamefont
  {Liu}}\ and\ \bibinfo {author} {\bibfnamefont {Gang}\ \bibnamefont {Chen}},\
  }\href@noop {} {\bibfield  {journal} {\bibinfo  {journal} {Unpublished}\ }
  (\bibinfo {year} {2018})}\BibitemShut {NoStop}%
\bibitem [{\citenamefont {Mydosh}\ and\ \citenamefont
  {Oppeneer}(2011)}]{RevModPhys.83.1301}%
  \BibitemOpen
  \bibfield  {author} {\bibinfo {author} {\bibfnamefont {J.~A.}\ \bibnamefont
  {Mydosh}}\ and\ \bibinfo {author} {\bibfnamefont {P.~M.}\ \bibnamefont
  {Oppeneer}},\ }\bibfield  {title} {\enquote {\bibinfo {title} {{Colloquium:
  Hidden order, superconductivity, and magnetism: The unsolved case of
  ${\mathrm{URu}}_{2}{\mathrm{Si}}_{2}$}},}\ }\href {\doibase
  10.1103/RevModPhys.83.1301} {\bibfield  {journal} {\bibinfo  {journal} {Rev.
  Mod. Phys.}\ }\textbf {\bibinfo {volume} {83}},\ \bibinfo {pages}
  {1301--1322} (\bibinfo {year} {2011})}\BibitemShut {NoStop}%
\bibitem [{\citenamefont {Li}\ \emph {et~al.}(2017{\natexlab{c}})\citenamefont
  {Li}, \citenamefont {Li}, \citenamefont {Yu}, \citenamefont {Paramekanti},\
  and\ \citenamefont {Chen}}]{Feiye_PRB2017}%
  \BibitemOpen
  \bibfield  {author} {\bibinfo {author} {\bibfnamefont {Fei-Ye}\ \bibnamefont
  {Li}}, \bibinfo {author} {\bibfnamefont {Yao-Dong}\ \bibnamefont {Li}},
  \bibinfo {author} {\bibfnamefont {Yue}\ \bibnamefont {Yu}}, \bibinfo {author}
  {\bibfnamefont {Arun}\ \bibnamefont {Paramekanti}}, \ and\ \bibinfo {author}
  {\bibfnamefont {Gang}\ \bibnamefont {Chen}},\ }\bibfield  {title} {\enquote
  {\bibinfo {title} {{Kitaev materials beyond iridates: Order by quantum
  disorder and Weyl magnons in rare-earth double perovskites}},}\ }\href
  {\doibase 10.1103/PhysRevB.95.085132} {\bibfield  {journal} {\bibinfo
  {journal} {Phys. Rev. B}\ }\textbf {\bibinfo {volume} {95}},\ \bibinfo
  {pages} {085132} (\bibinfo {year} {2017}{\natexlab{c}})}\BibitemShut
  {NoStop}%
\bibitem [{\citenamefont {Jackeli}\ and\ \citenamefont
  {Khaliullin}(2009)}]{Khaliullin}%
  \BibitemOpen
  \bibfield  {author} {\bibinfo {author} {\bibfnamefont {G.}~\bibnamefont
  {Jackeli}}\ and\ \bibinfo {author} {\bibfnamefont {G.}~\bibnamefont
  {Khaliullin}},\ }\bibfield  {title} {\enquote {\bibinfo {title} {{Mott
  Insulators in the Strong Spin-Orbit Coupling Limit: From Heisenberg to a
  Quantum Compass and Kitaev Models}},}\ }\href {\doibase
  10.1103/PhysRevLett.102.017205} {\bibfield  {journal} {\bibinfo  {journal}
  {Phys. Rev. Lett.}\ }\textbf {\bibinfo {volume} {102}},\ \bibinfo {pages}
  {017205} (\bibinfo {year} {2009})}\BibitemShut {NoStop}%
\bibitem [{\citenamefont {Liu}\ and\ \citenamefont
  {Khaliullin}(2018)}]{Khaliullin2}%
  \BibitemOpen
  \bibfield  {author} {\bibinfo {author} {\bibfnamefont {Huimei}\ \bibnamefont
  {Liu}}\ and\ \bibinfo {author} {\bibfnamefont {Giniyat}\ \bibnamefont
  {Khaliullin}},\ }\bibfield  {title} {\enquote {\bibinfo {title} {{Pseudospin
  exchange interactions in ${d}^{7}$ cobalt compounds: Possible realization of
  the Kitaev model}},}\ }\href {\doibase 10.1103/PhysRevB.97.014407} {\bibfield
   {journal} {\bibinfo  {journal} {Phys. Rev. B}\ }\textbf {\bibinfo {volume}
  {97}},\ \bibinfo {pages} {014407} (\bibinfo {year} {2018})}\BibitemShut
  {NoStop}%
\bibitem [{\citenamefont {Sano}\ \emph {et~al.}(2018)\citenamefont {Sano},
  \citenamefont {Kato},\ and\ \citenamefont {Motome}}]{Sano2}%
  \BibitemOpen
  \bibfield  {author} {\bibinfo {author} {\bibfnamefont {Ryoya}\ \bibnamefont
  {Sano}}, \bibinfo {author} {\bibfnamefont {Yasuyuki}\ \bibnamefont {Kato}}, \
  and\ \bibinfo {author} {\bibfnamefont {Yukitoshi}\ \bibnamefont {Motome}},\
  }\bibfield  {title} {\enquote {\bibinfo {title} {{Kitaev-Heisenberg
  Hamiltonian for high-spin ${d}^{7}$ Mott insulators}},}\ }\href {\doibase
  10.1103/PhysRevB.97.014408} {\bibfield  {journal} {\bibinfo  {journal} {Phys.
  Rev. B}\ }\textbf {\bibinfo {volume} {97}},\ \bibinfo {pages} {014408}
  (\bibinfo {year} {2018})}\BibitemShut {NoStop}%
\end{thebibliography}%

\end{document}